\documentclass[12pt]{article}
\usepackage[dvips]{graphicx}
\usepackage{amssymb}
\usepackage{amsmath}
\def\ii{{\rm i}}

\def\beq{\begin{equation}}\def\eeq{\end{equation}}
\def\bea{\begin{eqnarray}}\def\eea{\end{eqnarray}}

\begin{document}

\title{Expressing QFT in terms of QM with single extra dimension and classical hidden variable field}
\author{Roman Sverdlov,
\\Department of Mathematics, University of New Mexico} 

\date{June 24, 2020}
\maketitle

\begin{abstract}
The goal of this paper is to re-express QFT in terms of two "classical" fields living in ordinary space with single extra dimension. The role of the first classical field is to set up an injection from the set of values of extra dimension into the set of functions, and then said injection will be used in order to convert the second field into a coarse grained functional, thereby approximating QFT state. It turns out that this work also has a side-benefit of modeling ensemble of states in terms of one single state which, in turn, is interpretted in the above way. It is important to clarify that by "classical" we mean functions over ordinary space rather than configuration, Fock or function space. The "classical" theory that we propose is still non-local. 
\end{abstract}

\subsection*{1. Introduction}

It is generally assumed that the key problem with quantum mechanics is a problem with measurement. After all, apart from measurement, we have deterministic unitary evolution, while measurement outcome is random. Besides, in the absence of measurement, unitary evolution is well defined, while in case of measurement, there is no agreed upon definition as to what constitutes measurement as such. Finally, the unitary evolution can be viewed as local if we are concerned about field operators as opposed to actual states described over a hypersurface. On the other hand, measurement is distinctly non-local. 

While the above points are legitimate, I don't agree that they are the most crucial things that make quantum mechanics quantum. After all, some proposals have been made as to how to model the measurement: Bohmian mechanics (for example, \cite{Bohm1}), GRW model (see \cite{GRW1} and \cite{GRW2}), etc. While said proposals are non-local, Newtonian mechanics was non-local as well. Thus, our need for locality is due to \emph{empirical} evidence for relativity as opposed to what our classical intuition \emph{demands}. Therefore, as long as said theories claim to match conventional predictions -- which they do -- the empirical evidence can't falsify them, which is all we need. Perhaps more serious problem is that lack of falsification doesn't amount to a proof: after all, there is no agreement which of those several theories, if any, takes place in nature. That, again, is nothing new: people before Newton were facing these same problems, yet they weren't claiming that they should abandon classical logic. 

The consensus among conventional scientists is "impossibility" of reconciling quantum physics with classical intuition; but what we have here is not impossibility at all, just lack of knowledge. After all, if something is "impossible" then there is zero possible ways it could happen. On the other hand, if we "don't know enough" then there are multiple ways things could happen: the less we know the more there are possibilities that are consistent with our incomplete knowledge. This is exemplified in what happened in pre-Newtonian times with multiple theories of planet motion, and this is also what is happening with quantum mechanics today, with multiple competing theories of measurement. The fact that we have GRW competing with Bohm, as opposed to not having either one, is what points to the fact that our problem is lack of knowledge rather than impossibility. Once again, by looking at pre-Newtonian times, we see that lack of knowledge doesn't call for abandonment of classical logic. Thus, we don't have to do that in the context of quantum mechanics either. 

However, there is far more serious problem that \emph{does}, in fact, imply some sense of "impossibility", which is largely overlooked. In particular, the ontology of quantum state itself can't be viewed in the classical terms. And this is true even in the absence of measurement! Yes, unitary evolution is deterministic, but we don't know "what" said deterministic process is describing! 

But now we have to be a little more careful. If we talk about single particle quantum mechanics, we can easily answer the question we just posed by simply comparing Schr\"odinger's wave function to classical Maxwell field. Indeed, if we have no problem with Maxwell field changing directions, we shouldn't have any problem with $\psi$ being complex-valued. After all $\psi$ is not a probability, its a field.  The relation between probability and $\vert \psi \vert^2$ is similar to relationship between the probability and the weights placed on the two sides of biased coin. Said weights are still physical parameters, \emph{not} probabilities, and the same is true for $\psi$. 

The problem begins when we introduce multiparticle configuration space. In this case, Maxwell field is no longer a good analogy since it lives in ordinary space as opposed to configuration space. Perhaps this is what forces us to instead call $\psi$ a "probability amplitude" since "probability", in fact, lives in configuration space; but then the problem arises from the fact that probability as we know it is positive real, while $\psi$ is not. Keeping in mind all of the logical connections we just made, one can argue that the presence of configuration space is the single most important problem in quantum mechanics. Indeed, this point was named by various notable physicists (see \cite{Conf1} for some references). 

One might object to this by pointing out that in classical physics we also have configuration space. The important difference, though, is that in classical physics one is dealing with points in configuration space, while in quantum mechanics one is dealing with a wave function over a configuration space. A point in the configuration space can be reduced to a collection of points in ordinary space, while the wave function in a configuration space can not be reduced to the collection of wave functions in the ordinary space. There is, however, one example of a function over configuration space for which a different kind of reduction can be done. In particular, it is a positive valued probability density that is additive (which is the case classically but is not the case quantum mechanically). In this case, one can view such probability as \emph{epistomic} \cite{PBR}. Namely, it represents the state of knowledge of the observer. This allows for the possibility that if the observer's knowledge was complete, it would become $\delta$-function or, in other words, a single point in configuration space (and we already stated earlier that a point in a configuration space can be replaced by a collection of points in ordinary space). On the other hand,  in quantum mechanical case it is no longer additive: $\vert \psi_1+ \psi_2 \vert^2 \neq \vert \psi_1 \vert^2 + \vert \psi_2 \vert^2$. In this case, by PBR Theorem, $\psi$ is ontic \cite{PBR}, which means that $\psi$ corresponds to reality. This implies that $\psi$ is independent of observer's knowledge and, therefore, can not be reduced to $\delta$-function. This forces us to confront the issue of ontology of a wave function over the configuration space. 

Now, the difference between ordinary space and configuration space is simply that the latter has too many dimensions. Therefore, we address it by reducing that number. In case of $n$ particles, we replace $\mathbb{R}^{3n+1}$ by $\mathbb{R}^5$ by means of one single extra dimension, $y=x^5$. In our notation, $t=x^0$, $\vec{x} =(x^1, x^2, x^3)$, and $x=(t, \vec{x})=(x^0,x^1,x^2,x^3)$. We introduce space filling curve $\Gamma \colon y \mapsto \mathbb{R}^{3n}$. From that curve, we obtain $n$ different curves $\gamma_1, \cdots, \gamma_n$ where $\gamma_k (y)= (\Gamma_{3k-2}(y), \Gamma_{3k-1}(y), \Gamma_{3k}(y))$. We then picture $n$ different $1+1$ dimensional objects. The $1+1$ dimensional object number $k$ is a hypersurface spanned by the curve $\gamma_k$ that is stationary in $t$ (needless to say, quantum mechanics with fixed number of particles assumes Aristotelian spacetime). For any value $y=y_0$ we can intersect those hypersurfaces with a hypersurface $y=y_0$ to produce configuration of particles. Thid is the geometric interpretation of our space filling curve. Then the evolving quantum state, which is $\psi (t; \vec{x}_1, \cdots, \vec{x}_n)$ would correspond to $\xi (t,y)= \psi (t; \gamma_1 (y), \cdots, \gamma_n (y))$. But, in order to allign it with our classical intuition, we would like to replace $\xi (t,y)$ with $\xi (t, \vec{x}, y)$. This corresponds to an \emph{ensemble} of states parametrized by $\vec{x}$. Namely, $\xi (t,\vec{x},y)= \psi_{\vec{x}} (t; \gamma_1 (y), \cdots, \gamma_n (y))$. We physically understand this by adding an extra point particle, which we call a \emph{fly}. The state $\psi_{\vec{x}}$ represents a state where all the other articles are in the state $\psi$ and the fly is at a location $\vec{x}$. Just like with any other particle, we can put a fly into a state of fixed momentum by means of Fourier transform. That is, 
\begin{equation} \psi_{\vec{p}} (t; \gamma_1 (y), \cdots, \gamma_n (y)) = \frac{1}{(2 \pi)^{3/2}} \int d^3 x e^{i \vec{p} \cdot \vec{x}} \xi (t, \vec{x}, \gamma_1 (y), \cdots, \gamma_n (y)) \end{equation}
An ensemble of states can then be reduced to a single state that represents entanglement with the fly. Note that the correspondence will be different depending on whether we want to track position of the fly or its momentum. In one case states of the ensemble represent fixed position components, and in the other case they represent fixed momentum components. This implies that each state in one ensemble is an integral of the states in the other ensemple multiplied by appropriate exponent. The state involving the fly is one and the same, but the ensembles corresponding to that state will be different. 

Even though the constructions of space filling curves exist, they are not very physical. In order to make the theory more physical, we can \emph{drop} the assumption that $\Gamma$ is a space-filling curve and replace it with the assumption that $\Gamma$ hits $\epsilon$-neighborhood of most points in a certain bounded domain of configuration space, where $\epsilon$ is small but finite. One physical way in which this can happen is if we impose constraints $x^k \in [-\frac{L_k}{2}, \frac{L_k}{2}]$ and  $y \in [-\frac{L_5}{2}, \frac{L_5}{2}]$. If, at the same time, $L_5 \gg l_5 (\frac{L_1L_2L_3}{\epsilon^3})^n$ where $l_5$ is a length scale of random walk of these curves in $y$. Then,  one can reasonably expect a random curve $[- \frac{L_5}{2}, \frac{L_5}{2}] \mapsto ([- \frac{L_1}{2}, \frac{L_1}{2}] \times [- \frac{L_2}{2}, \frac{L_2}{2}]\times [- \frac{L_3}{2}, \frac{L_3}{2}])^n$ provided we make appropriate assumptions regatding the nature of the random walk in $y$ that would generate this curve. If we set $Y$ to infinity, then the curve would be ``onto up to infinitesimal". However, any given neighborhood of any given point will be hit infinitely many times and it would be difficult to write down the equations that would avoid the divergences. Instead, for the purposes of this paper, we will assume $Y$ is finite. But, at the same time, we also assume that $\epsilon$ is finite, so that the relation $L_5 \gg l_5 (\frac{L_1L_2L_3}{\epsilon^3})^n$ continues to hold. In this case, we would obtain a \emph{coarse grained} model of configuration space rather than the exact one. Any given state will be represented by \emph{infinitely many} choices of $\psi (\vec{x}_1, \cdots, \vec{x}_n)$ that are ``equivalent" to each other in the following sense: $\psi \sim \psi^{\prime}$ if and only if $\psi (\Gamma(y))= \psi^{\prime} (\Gamma (y))$ for all $y$.

Let is now apply this technique to QFT. In case of QFT, the configuration $\{\vec{x}_1, \cdots, \vec{x}_n\}$ gets replaced by $\phi (\vec{x})$ and, therefore, $\{\gamma_1(y), \cdots, \gamma_n(y)\}$ gets replaced with $\chi (\vec{x},y)$. Noticing that both $\phi (\vec{x})$ and $\chi (\vec{x},y)$ are $t$-independent, we conclude that we have to stick to the Aristotelian framework even in the case of QFT. The Lorentz invariance is accounted for by the fact that the Lagrangian -- which is introduced in Aristotelian framework -- \emph{happened} to be expressible in Lorentz invariant way. In order to understand the concept, picture a guitar string in Newtonian framework. An ant living on that string will be a believer in relativity with the ``wrong" speed of signals, which happens to be much slower than $c$. By the same token, $c$, itself, might be emergent as well. This means that, just like the ``true" speed of signal is much greater than the one on a guitar string, it is also much greater than $c$. In fact, it is infinite. That would justify modeling QFT within Aristotelian spacetime which, in turn, justifies the $t$-independence of $\phi (\vec{x})$ and $\chi (\vec{x},y)$ described earlier.

It is important to note that, once we adapt  $\chi (\vec{x},y)$, we have to reject $\{\gamma_1(y), \cdots, \gamma_n(y)\}$. This is because an $n$-particle QM state \emph{should} be obtained as a non-relativistic limit of QFT, yet $\{\gamma_1(y), \cdots, \gamma_n(y)\}$ does \emph{not} arise as a non-relativistic limit of $\chi (\vec{x},y)$. This can be traced down to the fact that a single particle corresponds to the first excited state of the harmonic oscillator, whose wave function has nothing to do with the $\delta$-function one might envision. Just like first excited state \emph{does} model a single particle, just not in the way we expected, there also \emph{is} a model of $n$-particle state within the model based on $\chi (\vec{x},y)$, it just wouldn't resemble our description based on $\gamma_1, \cdots, \gamma_n$. 

Since many body QM is merely a low energy limit of QFT, the "large number of dimensions" in the former case is a byproduct of infinitely many dimensions in the latter case. So we can restrict our quest to the dimensions present in QFT. Now, as far as QFT is concerned, it deals with harmonic oscillators in $\phi$. In QM case, the harmonic oscillator in $x$ corresponds to function $\psi (x)$. Therefore, in QFT case, the harmonic oscillator in $\phi$ corresponds to functional $\psi (\phi)$. Thus, the source of infinite dimensionality is simply that a functional is a function over infinite-dimensional domain. Therefore, in order to "get rid" of the problem, we have to replace functionals with ordinary functions. This is what I set out to do in this paper (and we focus exclusively on QFT since QM is merely its low energy limit). 

Clearly, the exact correspondence between functional and function is impossible for the simple reason that cardinalities are different. But, since there is no experimental proof that anything is exact, the approximate correspondence up to coarse graining would suffice us. In fact, the use of ultraviolet cutoff in QFT calculations implies that it is only defined up to certain scale anyway, it is simply that said scale happens to be very small and, therefore, unknown\footnote{Some people view cutoff as just a formalism and they won't make any conclusions based off of it, but our philosophy is to take things literally whenever possible so we do believe QFT has momentum upper bound, we simply don't know what it is.}.  This being the case, we propose to introduce a single extra coordinate, $y$, and use it as a way to parametrize subset of elements of $\phi$ that covers "enough" elements to "approximate" the QFT as we know it. This can be done by means of "hidden" classical field $\chi (\vec{x},y)$ which enables us to define $g^{(\chi)} \colon \{y \} \mapsto \{ \phi \}$ as 
\beq g^{(\chi)} (y) = \chi_y \eeq
where $\chi_y \colon \{ \vec{x} \} \mapsto \mathbb{R}$ is given by 
\beq \chi_y (\vec{x}) = \chi (\vec{x},y) \label{ChiyIntro} \eeq
This will enable us to replace $\psi (\phi)$ with $\xi (y)$. But, in order to have analogy with electromagnetic field, we would like to have $\xi (\vec{x},y)$ rather than $\xi (y)$. We do that by adding a non-interacting particle, which we call a "fly". Thus, we are describing all of the particles in a universe, plus a fly. If the particles in the universe are in a state that is conventionally represented by $\psi (\phi)$, and a fly has a momentum $\vec{p}$, then the function $\xi (\vec{x},y)$ takes the form
\beq \xi (\vec{x},y) = e^{i \vec{p} \cdot \vec{x}} \psi (g^{(\chi)} (y)) \eeq
Alternatively, we can utilize extra parameter as a way of defining ensemble of states as opposed to a single quantum state. Thus, the wave function
\beq \xi (\vec{x}, y) = \sum_k C_k e^{i \vec{p}_k \cdot \vec{x}} \psi_k (g^{(\chi)} (y)) \eeq
corresponds to the density matrix 
\beq \sum_k C_k \vert \psi_k (\phi) \rangle \langle \psi_k (\phi) \vert \eeq
What we are essentially saying is that, instead of ensemble of states, we have one single state that involves entanglement with a fly.  If the fly is non-interacting then the components of a state corresponding to different fly momenta will look like separate states in the ensemble. In reality they are part of one and the same state. This is certainly a good thing since some of the theories of quantum measurement (for example, quantum Darwinism) rely on the notion of ensemble of states which is another factor that takes away from realism, apart from the things talked about earlier. So it is good that we were able to address both question at the same time instead of making separate constructions for each one of them. This, in turn, will allow us to try convince realists to consider ensemble theories and conversely try to convince the ensemble people to consider realism.  

Going back to the issue of coarse graining, we have to warn the reader about the following problem. Suppose $R_{\vec{p}} (y) \in \mathbb{R}$ and $\Theta_{\vec{p}} (y)$ corresponds to amplitude and phase of the Fourier component of $\chi_y\colon \{ \vec{x} \} \mapsto \mathbb{R}$ (see Eq \ref{ChiyIntro}, \ref{RABC}, \ref{ThetaABC}). If we assume that $y$-coordinate is compactified, 
\beq y + L_5 = y \eeq
then the fact that $R_{\vec{p}}$ and $\Theta_{\vec{p}}$ are real valued implies that they are \emph{not} one to one. As far as $(R_{\vec{p}}, \Theta_{\vec{p}})$ is concerned, it \emph{might} be one to one, but it is not likely: after all it \emph{is} possible to draw a curve on a plane without self-intersections, yet a random curve is more likely to self-intersect than not. \emph{However}, if we consider \emph{three} parameters, $(R_{\vec{0}}, R_{\vec{p}}, \Theta_{\vec{p}})$, it \emph{is} most probably one to one: after all, the random curve in $\mathbb{R}^3$ is most likely \emph{not to} self intersect. If so, this creates a problem: any function $\xi \colon \{y \} \mapsto \mathbb{C}$, which we have \emph{intended} to correspond to the function over infinite dimensional domain, $\{ (R_{\vec{0}}, R_{\vec{p}_1}, \Theta_{\vec{p}_1}, R_{\vec{p}_2}, \Theta_{\vec{p}_2}, \cdots ) \} = \mathbb{R}^{\infty}$ can actually be modelled in terms of three dimensional domain, $\{ (R_{\vec{0}}, R_{\vec{p}}, \Theta_{\vec{p}} ) \}$. As will be explained later in more detail, $R_{\vec{0}}$ parameter can be used to model arbitrary number of particles with momentum $\vec{0}$, while $(R_{\vec{p}}, \Theta_{\vec{p}})$ can be used to model arbitrary number of particles with momenta $+ \vec{p}$ and $- \vec{p}$. Thus, an arbitrary state can be described as a linear combination of those three states! For example, 
\beq a^{\dagger}_{\vec{q}} \vert 0 \rangle = \sum_{abc} (a^{\dagger}_{\vec{p}})^a (a^{\dagger}_{- \vec{p}})^b (a^{\dagger}_{\vec{0}})^c \vert 0 \rangle \label{PhysicsContradiction} \eeq 
despite the fact that 
\beq \vec{p} \neq \vec{q} \label{pNeqq} \eeq 

In order to get out of this predicament, we make a claim that $a$-s, $b$-s and $c$-s on the right hand side will be forced to be extremely large numbers to the point of absurdity (in particular, the finer the coarse graining, the larger these numbers will have to be); the only choice of representation that avoids this feature is the one given on left hand side. The statement we just made might at first sound impossible: how can we isolate \emph{exactly one} state as opposed to narrow range of states? After all, the change of representation is continuous! Upon further look, however, there is no contradiction: we know that the set of basis states is discrete anyway; continuous change refers to the change in \emph{coefficients} next to afore-given set of basis states. Now, what we are saying is that if a coefficient of $(a_{\vec{p}}^{\dagger})^2 \vert 0 \rangle$ is to change by $0 (\epsilon)$, then the coefficient next to $(a_{\vec{p}}^{\dagger})^N \vert 0 \rangle$ will also change by $0 (\epsilon)$, for some $N \gg 1$. The continuity has nothing to do with $N \gg 1$; it has to do with $\epsilon \ll 1$, and the latter still holds. Now, it is conceivable that, due to some physical process, we would get the probability of $(a_{\vec{p}}^{\dagger})^N \vert 0 \rangle$ to be of $0 (\epsilon)$ rather than zero. The only thing we are trying to avoid is for that probability being large. Now, if we could change the probability of  $(a_{\vec{p}}^{\dagger})^N \vert 0 \rangle$ by $0 (\epsilon^2)$ while changing the probability of  $(a_{\vec{p}}^{\dagger})^2 \vert 0 \rangle$ by $0 (\epsilon)$, then $\epsilon^{-1}$ of those changes would lead to finite change of probability of  $(a_{\vec{p}}^{\dagger})^2 \vert 0 \rangle$ despite $0 (\epsilon)$ change of probability of  $(a_{\vec{p}}^{\dagger})^N \vert 0 \rangle$. But since in actuality both have change by $0 (\epsilon)$ at the same time, the above scenario is impossible. In other words, if we insist that $(a_{\vec{p}}^{\dagger})^N \vert 0 \rangle$ is of $0 (\epsilon)$ instead of large, then $(a_{\vec{p}}^{\dagger})^2 \vert 0 \rangle$ will have to be of $0 (\epsilon)$ rather than large, as well. Thus, we do have a narrow range of states instead of one single state, just as common sense tells us. And, indeed, we have to have narrow range of states on a physical grounds anyway, since we never know what traces of various past interactions could produce. 

Let us now go back to the statement we have made the next sentence after Eq \ref{pNeqq} and explain why we believe that statement. First of all, since the curve $g^{(\chi)} (y)$ fills the function space only up to coarse graining, it is impossible to shift in $(R_{\vec{q}}, \Theta_{\vec{q}})$ direction while keeping all the other $R$-s and $\Theta$-s constant. However, it is possible keep the latter \emph{approximately} constant: in particular, we have to "jump" by a very large distance in $y$ in such a way that, at the new point in $y$ the curve $g^{(\chi)} (y)$ "happens" to "come back to" the original point in projection to $(R_{\vec{p}}, \Theta_{\vec{p}})$, but not in projection to $(R_{\vec{q}}, \Theta_{\vec{q}})$. In other words, we change $(R_{\vec{q}}, \Theta_{\vec{q}})$ a lot while changing $(R_{\vec{p}}, \Theta_{\vec{p}})$ only slightly. Since at least one of those parameters changes a lot, $\xi (y)$ has to change a lot as well, there is no question about it. However, we can try to be silly and explain that change by the fact that $(R_{\vec{p}}, \Theta_{\vec{p}})$ had changed. In this case, the $(R_{\vec{p}}, \Theta_{\vec{p}})$-gradient of $\psi (\phi)$ better be very large since the change of $(R_{\vec{p}}, \Theta_{\vec{p}})$ is very small.  The only time when gradient of $\psi$ is large is when we are dealing with high energies. And since the momentum in question, $\vec{p}$, is fixed, the only way for energy to be large is to have large number of particles with that momentum -- which is where that statement is coming from. On the other hand, if we decide to be more reasonable and say that the cause of the change was $(R_{\vec{q}}, \Theta_{\vec{q}})$ after all, then we no longer need $\psi$ to change fast and therefore no longer need large number of particles. 

What we have said so far can be summarized as a trade-off between larger dimension and greater precision versus smaller dimension and lesser precision. On the one hand, one change cancels the other so both spaces are equal in size, which allows us to establish correspondence. On the other hand, we care about dimensionality a lot more than about precision, which is why "winning" the former is a huge accomplishment, even if it comes at the cost of "losing" the latter. From the field perspective, the above trade-off has to do with the fact we can choose $\xi (y)$ which is only one coordinate (thus making space smaller) yet can be measured precisely (thus making space larger), or we can choose $\psi (\phi)$ that has many degrees of freedom (thus making space larger) yet is only defined up to coarse graining (thus making space smaller). In the state language the tradeoff is that, on the one hand, we can impose the cutoff on the particle numbers (making space smaller) yet have many different momenta (making space larger) or we can have only three allowed momenta (making space smaller) yet allowing all particle numbers without any cutoffs (thus making space larger). 

The purpose of the rest of the paper is to make some of what we said a lot more explicit. We will do it in the following steps:

{\bf Sections 2:} We provide a model of many body QM in terms of a path integral of a single particle in $5$ dimensions. We will provide the model of Feynmann path integral based on the idea of coarse graining we described. 

{\bf Section 3:} We repeat what we did in Section 2 for the QFT case. This model will assume the Aristotelian spacetime yet the Lorentzian phenomenology will be conjectured to emerge due to our specific choice of Lagrangian. This model will, furthermore, be coarse grained. 

{\bf Section 4:} Start by writing down analytic solution for general excited state of harmonic oscillator in 1D and 2D. While in most textbooks one can look up the 1D solutions for the first few states (for example, \cite{Griffiths}), it is very difficult to find a book that will give the one for general excited state, much less its 2D version, so I decided to derive it myself to use it as a reference. Such derivation, however, turned out to be very long so I skipped most of it and only covered a brief summary of key steps.

{\bf Section 5:}  Convert the wave equations for general states of 1D and 2D oscillator into the equations of a functionals of general state. Similarly, convert the definitions of raising and lowering operator into the definitions of creation and annihilation operators by replacing ordinary derivatives with functional derivatives, coordinates with other functionals, and so forth.

{\bf Section 6:}  Use the ideas we talked about in order to replace $\psi (\phi)$ with $\xi (\vec{x},y)$, thus arriving with a definition of multiparticle state that "looks like" a single particle in 5D and, therefore, "realistic". We will likewise write down explicit expressions for creation and annihilation operators as well, which will include the need to define derivative in the context of coarse graining, and so forth. 

{\bf Section 7:} We describe the dynamics of "classical" field $\xi (\vec{x},y)$ which is something we haven't done in previous sections which are all focused on kinematical definitions of states. The goal of the dynamics proposed in Section 5 is to make sure that, if $\xi (\vec{x},t)$ obeys said "classical" (yet non-local) dynamics, then the corresponding quantum states (as defined in previous sections) will obey some version of coarse grained QFT. 

 Section 3 is to be contrasted to Sections 4-7 with the former being the counterpart to Feynmann path integral and the latter being the counterpart of Hamiltonian formalism. Since Sections 4-7 heavily rely on the Fourier transform, from the point of view of elegance and Ocams razor, perhaps the version of this paper consisting of just sections 1,3 and 8 (and skipping over 4-7) is the best. At the same time, however, we chose to include Sections 4-7 just to convince the skeptical reader that quantum states ``can" be described in our framework, as complicated as they might be. 

While the definition of general particle state will in fact be taken from Section 5 with appropriate modifications, the definition of creation and annihilation operators will be substantially different from Section 5 since in case of Section 5 we could use infinitesimal shifts while in case of Section 6 we couldn't. Since our goal is Section 6, we could have skipped the creation and annihilation operators in Section 5 if we wanted (the wave function of Section 5 was obtained by copying the one from Section 4, so we didn't need to write Section 5 version of creation and annihilation operators to derive it). The reason we included the definition of creation and annihilation operators in Section 5 is largely due to the wish for completeness. 

\subsection*{2. Realistic model of coarse grained path integral: QM toy model}

As stated earlier, QM should be viewed as a low energy limit of QFT but this does \emph{not} apply to the respective descriptions of QM and QFT we are proposing. After all we are using $\Gamma (y)$ for QM and $\chi (\vec{x},y)$ for QFT, which are different mathematical objects. It is true that their respective \emph{phenomenologies} obey this relation, but in this paper we are interested in ontology, not the phenomenology.  Therefore, our view is to stick to our QFT model and drop our QM model. Indeed the QM model will no longer be necessary since QM will arise out of QFT as a low energy limit. However, for \emph{pedagogical} purposes we will present a \emph{toy model} of what we \emph{could have done} if the QFT was not in the picture. Since the rest of the paper pertains to QFT, a reader is free to skip this sub-section. 

Consider fixed curves $\Gamma (y) = (\gamma_1 (y), \cdots, \gamma_n (y))$ where $\gamma_k$ are the functions of the form $ [- \frac{L_5}{2}, \frac{L_5}{2}] \mapsto  [- \frac{L_1}{2}, \frac{L_1}{2}] \times  [- \frac{L_2}{2}, \frac{L_2}{2}] \times  [- \frac{L_3}{2}, \frac{L_3}{2}]$. Assume $L_5 \gg l_5 (\frac{L_1L_2L_3}{\epsilon^3})^n$ where $l_5$ is the length factor of random walk in $x_5$ that generates $\Gamma (x_5)$ and design that random walk in such a way that it implies that it hits $\epsilon$-neighborhood of any given point in a configuration space. As explained earlier, we introduce an extra particle, which we call a ``fly" and we will describe an ensemble of states as a single state produced by the entanglement between various quantum states and the state of a ``fly": each state in the ensemble represents a component where the ``fly" is constrained to a given $\vec{x} \in \mathbb{R}^3$.

Our approach is to replace the path integral over $(\vec{x}_1, \cdots, \vec{x}_n)$ with the path integral over $y$, with the $\Gamma (y)$ building the bridge between the two. We will do that by picturing a fly that is feely moving along the $y$-axis, while its $\vec{x}$-coordinate is fixed. Furthermore, we will utilize $\Gamma$ in order to define the distance along $y$-axis:
\begin{equation} d(y_1,y_2)= \bigg(\sum_{k=1}^n \vert \gamma_k (y_1)- \gamma_k (y_2) \vert^2 \bigg)^{1/2} \end{equation}
we also define a potential 
\begin{equation} U(y)= V(\gamma_1 (y), \cdots, \gamma_n (y)) \end{equation}
where $V$ is the agreed upon potential on a configuration space. From the point of view of ontology, we have $V(\vec{x},y)$ rather than $V(y)$. It is just that $V(\vec{x}, y)$ is $\vec{x}$-independent, 
\begin{equation} V(\vec{x},y)= V(y) \end{equation}
We then use path integral intuition to write
\begin{equation} \xi (t-\delta t, \vec{x},y) =\int dy^{\prime} \xi (t, \vec{x},y^{\prime}) \exp\bigg[ i \bigg(\bigg(\frac{d (y,y^{\prime})}{\delta t} \bigg)^2 - V(y) \bigg) \bigg] dy^{\prime} \end{equation}
which, upon the substitution of the $V$ and $d(y,y^{\prime})$ becomes
\begin{equation} \xi (t- \delta t, \vec{x},y) =\int dy^{\prime} \xi (t, \vec{x},y^{\prime}) \exp\bigg[ i \bigg(\sum_{k=1}^n \frac{\vert \gamma_k (y_1)- \gamma_k (y_2) \vert^2}{(\delta t)^2} -  V(\gamma_1 (y), \cdots, \gamma_n (y))  \bigg) \bigg] dy^{\prime} \end{equation}

\subsection*{3. Realistic model of coarse grained path integral: QFT case}

Consider the situation where the ``fly" is constrained to the fixed coordinates $\vec{x}$ and is freely moving along $y$-axis. Suppose it's path is a zigzag-type. It takes the time $\delta t$ for the ``fly" to make a jump, which goes along the straight line. The ``jumps" occur one after the other: if the first ``jump" started at $t_0 - \delta t$ and ended on $t_0$ then the next ``jump" starts from $t_0$ and ends on $t_0 + \delta t$. The end-point of the previous jump is the same as the starting point of the next jump. However, the directions of these jumps are independent of each other. By this we mean that the end-point of the next jump is only a function of the starting point of that jump (which happen to coincide with the end-point of the previous jump) -- but it is independent of the starting point of the previous jump. 

On the other hand, in place of ``length" of the jump we have to put $\vert \chi_{y_2} - \chi_{y_1} \vert$ as opposed to $\vert y_2- y_1 \vert$. We put ``length" in quotes because we affirm that the \emph{actual} length is $\vert y_2 -y_1 \vert$: we have to, or else we ruin the main purpose of what we are doing in this paper (restoration of our geometric intuition). But, at the same time, we have to put $\vert y_2- y_1 \vert$ into our formula in order for the dynamics to approximate the well known one. The square of $\vert \chi_{y_2} - \chi_{y_1} \vert$ is given by 
\begin{equation}\vert \chi_{y_2} - \chi_{y_1} \vert^2 = \int dx^{\prime} (\chi (x^{\prime}, y^{\prime}) - \chi (x^{\prime},y))^2 \end{equation}
We also have to include the ``potential" term in path integral. Clearly, in $\phi^4$-theory, the $\phi^4$ term is part of the ``potential". But its not the only part. Apart from that, the ``mass" term $\frac{m^2}{2} \phi^2$ is also part of the potential and, on a more controversial note, $\vert \vec{\nabla} \phi \vert^2$ is part of the potential, as well. The fact that $\vert \vec{\nabla} \phi \vert^2$ is part of the potential while $(\partial \phi/ \partial t)^2$ isn't, underscores the fact that our spacetime is Aristotelian as opposed to Lorentzian. The Lorentzian phenomenology results from the specific structure of Lagrangian; namely, that Lagrangian happens to be written in Lorentz invariant form. 

Finally, we are going to do a little modification to what we said. We said that the ``fly" is undergoing zigzag path. However, since we are ``quantizing" that fly, we do not have to assume a specific partition of time into $\delta t$ intervals. Instead, we can write a continuum equation for $\partial \xi/ \partial t$, which is written \emph{as if} the time interval happened to end at that particular $t$. In order to understand this, note that if we did consider a fixed partition of time into the intervals, we would be summing over different zigzags. This would include zigzags that are time-shifted by $n \delta t$ relative to other zigzags. This being the case, their \emph{superposition} won't change its behavior that much within a small number of $\delta t$-intervals, despite the fact that each \emph{particular} zigzag would. Since we are interested in superposition of zigzags as opposed to each specific one, we ``can afford" to err by a few $\delta t$-intervals. But, if so, we can ``even more" afford to err by a fraction of $\delta t$-interval by pretending that it ends at the point $t$ even if it does not. This leads us to write
\begin{equation} \xi(t, \vec{x},y) = \int dy^{\prime} \xi(t, \vec{x},y^{\prime}) \exp \bigg[i \bigg( \int dx^{\prime} \bigg(\frac{\chi (x^{\prime}, y^{\prime}) - \chi (x^{\prime}, y)}{\delta t}\bigg)^2 - V(\chi, \vec{\nabla}_x \chi;x^{\prime},y) \bigg) \bigg]  \label{SimplerDynamics} \end{equation}
This equation leads to \emph{continuous} and \emph{differentiable} behavior of $\xi$, similar to the continuous and differentiable behavior of $\psi$ in first quantization. Thus, from the ontological point of view, we have a first quantization in 5D, which is meant to represent the second quantization in 4D, just like the title of this paper suggests. 

What we said so far, in principle, has accomplished our goal of expressing QFT in 4 dimensions in terms of QM in 5 dimensions. However, in order to convince the reader this is the case, let us explicitly translate the key concepts pertaining to 4-dimensional Fock space into the context of 5-dimensional QM. These translations will be very awkward and lengthy. But the reader should keep in mind that they do not have to do with physics as such: the physics itself is complete with Chapter 2 alone. The one exception to this statement is Chapter 6 where we \emph{will} return to physics. But the physics given in Chapter 6 is only hypothetical. What we mean by this is that the dynamics has to be \emph{either} Chapter 2 \emph{or} Chapter 6 -- not both. For completeness purposes, we are offering the reader a choice between the two -- even though we, ourselves, favor Chapter 2. With this prelude, let us now switch to the aforementioned ``awkward constructions" that would translate Fock space into our framework (Chapters 3-5). 

\subsection*{4. Review of harmonic oscillator in 1D and 2D}

Throughout the rest of the paper, we will be using the results of the wave function of general excited state of 2D harmonic oscillator presented in \cite{viXra}. Since deriving general state (as opposed to first few excited states) is quite difficult, there is no way we could repeat the adequate derivation here. Therefore, we will briefly summarize the few key steps, and the reader is referred to \cite{viXra} for more detail. 

Creation and annihilation operators are defined as
\beq a^{\dagger} = \sqrt{\frac{m \omega}{2}} x - \frac{1}{\sqrt{2m \omega}} \frac{d}{dx} \; , \; a = \sqrt{\frac{m \omega}{2}} x + \frac{1}{\sqrt{2m \omega}} \frac{d}{dx} \eeq
and satisfy commutation relations
\beq [a, a^{\dagger}]=1 \eeq
The first three states are 
\beq \psi_0 (x) = \bigg(\frac{m \omega}{\pi} \bigg)^{1/4} e^{-m \omega x^2/2} \label{1D0} \eeq
\beq \psi_1 (x) = \frac{2^{1/2} (m \omega)^{3/4}}{\pi^{1/4}} x e^{- m \omega x^2/2} \label{1D1} \eeq
\beq \psi_2 (x) = \bigg(\frac{m \omega}{\pi} \bigg)^{1/4} e^{-m \omega x^2/2} \bigg( \sqrt{2} m \omega x^2 - \frac{1}{\sqrt{2}} \bigg) \label{1D2} \eeq
One can see by induction that $n$-th excited state can be expressed as 
\beq \psi_n (x) = \frac{1}{\sqrt{n!}} \bigg( \frac{m \omega}{\pi} \bigg)^{1/4}  \bigg( \sqrt{\frac{m \omega}{2}} \hat{x} - \frac{1}{\sqrt{2m \omega}} \frac{d}{dx}  \bigg)^n e^{-m \omega \hat{x}^2/2}   \label{InductionStart1D} \eeq
which can be rewritten as 
\beq \psi_n (x) = \frac{1}{\sqrt{n!}} \bigg( \frac{m \omega}{\pi} \bigg)^{1/4} e^{-m \omega \hat{x}^2/2} \bigg[ e^{m \omega \hat{x}^2/2} \bigg( \sqrt{\frac{m \omega}{2}} \hat{x} - \frac{1}{\sqrt{2m \omega}} \frac{d}{dx} \bigg) e^{-m \omega \hat{x}^2/2} \bigg]^n 1 \eeq
and then further rewritten as 
\beq \psi_n (x) = \frac{1}{\sqrt{n!}} \bigg( \frac{m \omega}{\pi} \bigg)^{1/4} e^{-m \omega \hat{x}^2/2} \bigg( \sqrt{2 m \omega} \hat{x} - \frac{1}{\sqrt{2m \omega}} \frac{d}{dx} \bigg)^n 1 \eeq 
to obtain, after some combinatorics,
\beq \psi_n (x) = \sqrt{n!} \bigg( \frac{m \omega}{\pi} \bigg)^{1/4} e^{-m \omega x^2/2} \sum_{C=0}^{\lfloor n/2 \rfloor} \frac{(-1)^C (2m \omega)^{\frac{n}{2} - C}}{2^C C! (n-2C)!} x^{n-2C} \label{nthState1D} \eeq
Now, the factor $1/\sqrt{n!}$ in Eq \ref{InductionStart1D} was specifically designed in such a way that each state in the ladder is properly normalized. Yet, the normalization of Eq \ref{nthState1D} is not at all obvious. It turns out, however, that the normalization follows from the following identity, 
\beq p+q \; {\rm is \; even} \; \Longrightarrow \label{norm1D6} \eeq
\beq \Longrightarrow \; \sum_{c_1=0}^{\lfloor p/2 \rfloor} \sum_{c_1=0}^{\lfloor q/2 \rfloor}\bigg( \frac{(-1)^{c_1+c_2}}{ c_1! c_2! (p-2c_1)! (q-2c_2)!}   \frac{(p+q-2c_1-2c_2)! }{ (\frac{p+q}{2}-c_1-c_2)! } \bigg) = \frac{2^p}{p!} \delta^p_q= \frac{2^q}{q!} \delta^p_q \nonumber \eeq
which I have proven in the separate paper that I am working on getting published, but it would be too much of a sidetrack to include that proof here. 

In two dimensional case, the harmonic oscillator has two degrees of freedom. In Cartesian coordinates these would be coming from oscillators in either of the two axes, and in polar coordinates these would be coming from total energy and angular momentum. Within these two degrees of freedom we define the operators
\beq a_{++}=  \frac{a_x^{\dagger} +i a_y^{\dagger}}{\sqrt{2}} \; , \; a_{+-}= \frac{a_x^{\dagger} - ia_y^{\dagger}}{\sqrt{2}} \label{a++a+-} \eeq
\beq a_{-+} = \frac{a_x+ia_y}{\sqrt{2}} \; , \; a_{--}= \frac{a_x-ia_y}{\sqrt{2}} \label{a-+a--} \eeq
We chose the above notation in such a way that the first sign represents what happens to energy upon action of said operator and the second sign represents what happens to angular momentum. Thus, $a_{++}$ raises both energy and angular momentum, $a_{--}$ lowers both, $a_{+-}$ raises energy while lowering angular momentum and $a_{-+}$ lowers energy while raising angular momentum. The Hermitian conjugate merely permutes those operators via the following expressions: 
\beq a_{++}^{\dagger} = a_{--} \; , \; a_{+-}^{\dagger} = a_{-+} \; ,\; a_{-+}^{\dagger} = a_{+-} \; , \; a_{--}^{\dagger} = a_{++} \eeq
and also the operators satisfy the following commutation relations:
\beq [a_{-+},a_{+-}]=[a_{--},a_{++}] = 1 \label{2DComm1} \eeq
\beq [a_{+-},a_{-+}]=[a_{++},a_{--}]=-1 \label{2DComm2} \eeq
\beq [a_{++},a_{-+}]=[a_{-+},a_{++}]= [a_{+-},a_{--}]= [a_{--},a_{+-}]=0 \label{2DComm3} \eeq
\beq [a_{++},a_{+-}]=[a_{+-},a_{++}]=[a_{-+},a_{--}]=[a_{--},a_{-+}]=0 \label{2DComm4} \eeq
\beq [a_{++},a_{++}]=[a_{+-},a_{+-}]=[a_{-+},a_{-+}]=[a_{--},a_{--}]=0 \label{2DComm5} \eeq
In polar coordinates those operators are defined as
\beq a_{++} =  \frac{e^{i \theta}}{2} \bigg( r \sqrt{m \omega} - \frac{1}{\sqrt{m \omega}} \frac{\partial}{\partial r} - \frac{i}{r \sqrt{m \omega}} \frac{\partial}{\partial \theta} \bigg) \label{128b} \eeq
\beq a_{+-} =  \frac{e^{-i \theta}}{2} \bigg( r \sqrt{m \omega} - \frac{1}{\sqrt{m \omega}} \frac{\partial}{\partial r} + \frac{i}{r \sqrt{m \omega}} \frac{\partial}{\partial \theta} \bigg) \label{128d} \eeq
\beq a_{-+} = \frac{e^{i \theta}}{2} \bigg( r \sqrt{m \omega} + \frac{1}{\sqrt{m \omega}} \frac{\partial}{\partial r} + \frac{i}{r \sqrt{m \omega}} \frac{\partial}{\partial \theta} \bigg) \label{128a} \eeq
\beq a_{--} = \frac{e^{-i \theta}}{2} \bigg( r \sqrt{m \omega} + \frac{1}{\sqrt{m \omega}} \frac{\partial}{\partial r} - \frac{i}{r \sqrt{m \omega}} \frac{\partial}{\partial \theta} \bigg) \label{128c} \eeq
The ground state, $\psi_{00}$, has energy $1/2 + 1/2 = 1$ (coming from the oscillator in $x$ direction and another oscillator in $y$ direction) and angular momentum $0$; its wave function is
\beq \psi_{00} (r, \theta) = \sqrt{\frac{m \omega}{\pi}} e^{-m \omega r^2/2} \label{2D0} \eeq
Unlike 1D oscillator, there are two "first excited states", each having energy $1+1=2$. In Cartesian coordinates they are $a_x^{\dagger} \vert 0 \rangle$ and $a_y^{\dagger} \vert 0 \rangle$, corresponding to wave functions $\psi_1 (x) \psi_0 (y)$ and $\psi_0 (x) \psi_1 (y)$, while in polar coordinates they are $a_{++} \vert 0 \rangle$ and $a_{+-} \vert 0 \rangle$, corresponding to wave functions $\psi_{1,1} (r, \theta)$ and $\psi_{1,-1} (r, \theta)$ (where $\psi_{nL}$ denotes the state of $n$-th energy level (or, eqivalently, an energy of $n+1$) and angular momentum $L$). Each of the first pair of states can be represented as a linear combination of second pair of states, and visa versa; the energy of all four states is $2$. In polar coordinates, the state with energy $2$ and angular momentum $1$ is 
\beq \psi_{1,-1} (r, \theta) = \frac{m \omega}{\sqrt{\pi}} r e^{-m \omega r^2/2} e^{-i \theta} \label{2D1-1} \eeq
and the state with energy $2$ and angular momentum $-1$ is 
\beq \psi_{1,1} (r, \theta) = \frac{m \omega}{\sqrt{\pi}} r e^{-m \omega r^2/2} e^{i \theta} \label{2D11} \eeq
The fact that 
\beq \frac{m \omega}{\sqrt{\pi}} r e^{-i \theta} e^{-m \omega r^2/2}  = \frac{m \omega}{\sqrt{\pi}} (r \cos \theta - i r \sin \theta) e^{-m \omega (x^2 +y^2)/2} = \eeq
\beq = \frac{m \omega}{\sqrt{\pi}} (x - i y) e^{-m \omega x^2/2} e^{-m \omega y^2/2} = \frac{m \omega}{\sqrt{\pi}} \Big( \Big(x e^{-m \omega x^2/2} \Big) \Big( e^{-m \omega y^2/2} \Big) - i \Big(  e^{-m \omega x^2/2} \Big) \Big( y e^{-m \omega y^2/2} \Big) \Big) \nonumber \eeq 
and also that 
\beq \frac{m \omega}{\sqrt{\pi}} r  e^{i \theta} e^{-m \omega r^2/2}= \frac{m \omega}{\sqrt{\pi}} (r \cos \theta + i r \sin \theta) e^{-m \omega(x^2  + y^2 )/2} = \eeq
\beq = \frac{m \omega}{\sqrt{\pi}} (x + i y) e^{-m \omega x^2/2} e^{-m \omega y^2/2} = \frac{m \omega}{\sqrt{\pi}} \Big( \Big(x e^{-m \omega x^2/2} \Big) \Big( e^{-m \omega y^2/2} \Big) + i \Big(  e^{-m \omega x^2/2} \Big) \Big( y e^{-m \omega y^2/2} \Big) \Big) \nonumber \eeq 
confirms that, indeed, the two excited states in polar coordinates are linear combinations of the two excited states in Cartesian coordinates. 

Similarly, there are three "second excited states", with energy $1+2=3$. In Cartesian coordinates these are $a_x^{\dagger} a_x^{\dagger} \vert 0 \rangle$, $a_x^{\dagger} a_y^{\dagger} \vert 0 \rangle$ and $a_y^{\dagger} a_y^{\dagger} \vert 0 \rangle$, corresponding to wave functions $\psi_2 (x) \psi_0 (y)$, $\psi_1 (x) \psi_1 (y)$ and $\psi_0 (x) \psi_2 (y)$ (the reason we skipped $a_y^{\dagger} a_x^{\dagger} \vert 0 \rangle$ is that $[a_x^{\dagger}, a_y^{\dagger}]=0$). In polar coordinates, the three second excited states are $a_{++} a_{++} \vert 0 \rangle$, $a_{++} a_{+-} \vert 0 \rangle$ and $a_{+-} a_{+-} \vert 0 \rangle$, corresponding to wave functions $\psi_{22} (r, \theta)$, $\psi_{20} (r, \theta)$ and $\psi_{2,-2} (r, \theta)$ (once again, we skipped $a_{-+} a_{++} \vert 0 \rangle$ because $[a_{-+}, a_{++} ] =0$). The polar coordinate wave functions are given by
\beq \psi_{2,-2} = \frac{(m \omega)^{3/2}}{\sqrt{2 \pi}} r^2 e^{-2 i \theta} e^{-m \omega r^2/2} \label{2D2-2} \eeq
\beq \psi_{20} = \bigg( \frac{(m \omega)^{3/2}}{\sqrt{\pi}} r^2 - \sqrt{\frac{m \omega}{\pi}} \bigg) e^{-m \omega r^2/2} \label{2D20} \eeq
\beq \psi_{2,2} = \frac{(m \omega)^{3/2}}{\sqrt{2 \pi}} r^2 e^{2 i \theta} e^{-m \omega r^2/2} \label{2D22} \eeq
It is easy to see that 
\beq r^2 e^{2 i \theta} = r^2 (\cos 2 \theta + i \sin 2 \theta) = r^2 (\cos^2 \theta - \sin^2 \theta + 2 i \sin \theta \cos \theta) = \nonumber \eeq
\beq = (r \cos \theta)^2 + (r \sin \theta)^2 - 2 i (r \cos \theta) (r \sin \theta) = x^2 - y^2 -2ixy \eeq 
and, similarly,
\beq r^2 e^{-2 i \theta} = x^2-y^2+2ixy \eeq
Therefore, 
\beq \psi_{2,-2} = \frac{(m \omega)^{3/2}}{\sqrt{2 \pi}} (x^2+y^2+2ixy) e^{-m \omega (x^2+y^2)/2} =  \nonumber \eeq
\beq = \frac{(m \omega)^{3/2}}{\sqrt{2 \pi}} \bigg(\bigg(x^2 e^{-m \omega x^2/2} \bigg) \bigg( e^{-m \omega y^2/2} \bigg)+ \eeq
\beq +  \bigg( e^{-m \omega x^2/2} \bigg) \bigg(y^2 e^{-m \omega y^2/2}\bigg) + 2i \bigg(x e^{-m \omega x^2/2} \bigg) \bigg(y e^{-m \omega y^2/2} \bigg) \bigg) \nonumber \eeq
and  
\beq \psi_{2,2} = \frac{(m \omega)^{3/2}}{\sqrt{2 \pi}} (x^2+y^2-2ixy) e^{-m \omega (x^2+y^2)/2} =  \nonumber \eeq
\beq = \frac{(m \omega)^{3/2}}{\sqrt{2 \pi}} \bigg(\bigg(x^2 e^{-m \omega x^2/2} \bigg) \bigg( e^{-m \omega y^2/2} \bigg)+ \eeq
\beq +  \bigg( e^{-m \omega x^2/2} \bigg) \bigg(y^2 e^{-m \omega y^2/2}\bigg) - 2i \bigg(x e^{-m \omega x^2/2} \bigg) \bigg(y e^{-m \omega y^2/2} \bigg) \bigg) \nonumber \eeq
Finally, 
\beq \psi_{20} = \bigg( \frac{(m \omega)^{3/2}}{\sqrt{\pi}}(x^2+y^2) - \sqrt{\frac{m \omega}{\pi}} \bigg) e^{-m \omega (x^2+y^2)/2} = \nonumber \eeq
\beq = \frac{(m \omega)^{3/2}}{\sqrt{\pi}}  \bigg(x^2 e^{-m \omega x^2/2} \bigg) \bigg( e^{-m \omega y^2/2} \bigg) + \eeq
\beq + \frac{(m \omega)^{3/2}}{\sqrt{\pi}} \bigg(e^{-m \omega x^2/2} \bigg) \bigg(y^2 e^{-m \omega y^2/2} \bigg)  - \sqrt{\frac{m \omega}{\pi}} \bigg(e^{-m \omega x^2/2} \bigg) \bigg(e^{-m \omega y^2/2} \bigg) \nonumber \eeq 
This, indeed, confirms that any given state in polar coordinates can be represented as a linear combination of products of Cartesian coordinate states. 

The derivation of general state in polar coordinates would be too long of a sidetrack as far as this paper is concerned (although another paper with that derivation is in preparation). But let me give you a basic outline of steps that would serve as a brief summary of otherwise lengthy derivation. First, one can use Eq \ref{nthState1D} to write down $\psi_n (x) \psi_0 (y)$ as 
\beq \psi_n (x) \psi_0 (y) = \sum_{0 \leq k \leq n \; {\rm and} \; n-k \; {\rm is \; even}}^n \alpha_k x^k e^{-m \omega (x^2+y^2)/2} \label{CartesianStart} \eeq
then one can use 
\beq x^k = (r \cos \theta)^k = r^k \bigg(\frac{e^{i \theta}+ e^{-i \theta}}{2} \bigg)^k = \bigg(\frac{r}{2} \bigg)^k \sum_{l \in \{-k, -k+2, \cdots, k-2, k \}} {k \choose (n+l)/2} e^{il \theta} \eeq
to rewrite it as 
\beq \psi_n (x) \psi_0 (y) = \nonumber \eeq
\beq = \sum_{l \in \{-n, -n+2, \cdots, n-2, n \}} \bigg( e^{il \theta} \sum_{k \in \{-n, -n+2, \cdots, -l-2, -l \} \cup \{l, l+2, \cdots, n-2, n \}} \alpha_k \bigg(\frac{r}{2} \bigg)^k  {k \choose (n+l)/2} \bigg) \label{DoubleSumToy} \eeq
Then by noticing that 
\beq \hat{H} (\psi_n (x)\psi_0 (y)) = \bigg(n + \frac{1}{2} \bigg) \psi_n (x) \psi_0 (y) + \frac{1}{2} \psi_n (x) \psi_0 (y) = (n+1) \psi_n (x) \psi_0 (y) \eeq
\beq \hat{L} = \hat{x} \hat{p}_y - \hat{y} \hat{p}_x = - i \partial_{\theta} \eeq
one can deduce that the state with energy $n+1$ and angular momentum $L$ is $e^{iL \theta}$-term in Eq \ref{DoubleSumToy} up to some normalization constant; namely, 
\beq \psi_{nL} (r, \theta) =   N_{nL} e^{il \theta} \sum_{k \in \{-n, -n+2, \cdots, -l-2, -l \} \cup \{l, l+2, \cdots, n-2, n \}} \alpha_k \bigg(\frac{r}{2} \bigg)^k  {k \choose (n+l)/2}  \eeq
where $N_{nL}$ is the normalization coefficient. In order to find $N_{nL}$, we first find $N_{nn}$ (corresponding to $L=n$) since it turns out the easiest one to find and, afterwords, we see how that coefficient changes upon action of $a_{+-}$. Since $a_{+-}$ raises energy and lowers angular momentum, we anticipate to see $n$ replaced with $n+1$ and $l$ with $l-1$, thus obtaining
\beq a_{+-} \bigg(  e^{il \theta} \sum_{k \in \{-n, -n+2, \cdots, -l-2, -l \} \cup \{l, l+2, \cdots, n-2, n \}} \alpha_k \bigg(\frac{r}{2} \bigg)^k  {k \choose (n+l)/2} \bigg) = \nonumber \eeq
\beq = M_{nL}  e^{i(l-1) \theta} \sum_{k \in \{-n-1, -n+1, \cdots, -l-1, -l+1 \} \cup \{l-1, l+1, \cdots, n-1, n+1 \}} \alpha_k \bigg(\frac{r}{2} \bigg)^k  {k \choose (n+l)/2} \label{PolarIsPreserved} \eeq
However, we will have to perform explicit calculation in order to see what $M_{nL}$ is (said calculation is performed with $a_{+-}$ being expressed in polar coordinates). After finding out $M_{nL}$, we rewrite it as 
\beq a_{+-} \frac{\vert \psi_{nL} \rangle}{N_{nL}} = M_{nL} \frac{\vert \psi_{n+1,L-1} \rangle}{N_{n+1,L-1}} \eeq
and, in combination with 
\beq [a_{+-}, a_{-+}]=1 \eeq
as well as the value of $N_{nn}$, we find by induction the value of $N_{n+j, n-j}$, and, therefore, $N_{nL}$. As stated earlier, the explicit derivation is a lot lengthier than what is presented (in particular, the coefficients $\alpha_k$ need to be explicit, and so forth). After said derivation is done, the final answer will be 
\beq \psi_{nL} (r, \theta) =  \sqrt{\frac{ m \omega}{\pi}} \sqrt{2^n \Big(\frac{n-L}{2} \Big)! \Big(\frac{n+L}{2} \Big)!}  e^{-m \omega r^2/2}   \times \nonumber \eeq
\beq \times   \sum_{C=0}^{\min \big( \frac{n-L}{2}, \frac{n+L}{2} \big)} \frac{(-1)^C (2m \omega)^{\frac{n}{2} - C} r^{n-2C} e^{iL \theta}}{2^{n-C} C! \big( \frac{n+L}{2}-C \big)! \big(\frac{n-L}{2} -C \big)!}   \label{nLState2D} \eeq
If you check the normalization of Eq \ref{nLState2D}, the result might not look right: instead of $1$ you would get a rather complicated sum. However, you will find that the identity 
\beq  p+q \; {\rm is \; even} \; \Longrightarrow \nonumber \eeq
\beq \Longrightarrow \sum_{c_1=0}^{\min \big( \frac{p-L}{2}, \frac{p+L}{2} \big)}  \sum_{c_2=0}^{\min \big( \frac{q-L}{2}, \frac{q+L}{2} \big)} \frac{(-1)^{c_1+c_2} (\frac{p+q}{2}-c_1-c_2)! }{c_1! c_2! \big(\frac{p+L}{2}-c_1 \big)! \big(\frac{p-L}{2}-c_1 \big)!  \big(\frac{q+L}{2}-c_2 \big)! \big(\frac{q-L}{2}-c_2 \big)!}  = \nonumber \eeq
\beq = \frac{\delta^p_q}{ \sqrt{\big(\frac{p-L}{2} \big)! \big(\frac{p+L}{2} \big)! \big(\frac{q-L}{2} \big)! \big(\frac{q+L}{2} \big)!}}  \label{DesiredSumnL}  \eeq
implies that the normalization is as desired. The reader can check numerically that, indeed, the above identity holds. I have also written analytic proof of it, but that would be too much of a sidetrack for this paper so I will publish that proof separately. 

Clearly, there is another way of doing it. Instead of starting out from Cartesian coordinates in Eq \ref{CartesianStart}, one could have started from the ground state $e^{-m \omega r^2/2}$ and then work one's way up with $a_{++}$ and $a_{+-}$ by exclusively using polar coordinates. As Eq \ref{PolarIsPreserved} indicates, one would also arrive at Eq \ref{nLState2D} at the end of the day. The only problem with this approach is that Eq \ref{nLState2D} is very hard to guess by merely looking at the first few excited states -- unless one somehow anticipates that equation ahead of time. And the way to "anticipate" it is to start out from Cartesian coordinates as we have illustrated. 

Going back to "Cartesian coordinate start", one could have started from $\psi_{n_1} (x) \psi_{n_2} (y)$ instead of $\psi_n (x) \psi_0 (y)$.  However, the inspection of the above steps shows that the polar coordinate states derived from $\psi_n (x) \psi_0 (y)$ are just as general as the ones derived from $\psi_{n_1} (x) \psi_{n_2} (y)$. After all, $\psi_n (x) \psi_0 (y)$ "covers" all possible $\vert L \vert \leq n$  and if one then "runs" through all possible $n$-s one can see that we indeed "cover" all possible states (since none of the states with $\vert L \vert > n$ are allowed). So, since both $\psi_n (x) \psi_0 (y)$ and $\psi_{n_1} (x) \psi_{n_2} (y)$ result in equally general states yet the latter involves far more complicated calculations than the former, the $\psi_n (x) \psi_0 (y)$ approach is preferred. One could, however, still do $\psi_{n_1} (x) \psi_{n_2} (y)$ just to check that no mistakes were made. But if one looks harder one can see a long list of other things one might want to check which would lead to equally difficult calculations. At the end of the day one should simply trust that said calculations would go through. 

In order to see some of the verifications of how some results match, as well as more details of the deriving the formulae presented, the reader is referred to \cite{viXra}.

\subsection*{5. Representing QFT states as functionals}

In quantum mechanics case, the harmonic oscillator can be viewed as either a wave function $\psi (x)$ or as a linear combination of states defined via ladder operators. Now, a generic QFT state is defined in terms of the latter, where ladder operators are replaced with creation and annihilation operators. Thus, logic tells us, that said state can also be described as $\psi (\phi)$. Here, we have replaced $x$ with $\phi$ since in QM case Hamiltonian is a function of $x$ while in QFT it is a function of $\phi$. In other words, QFT state should be described as a functional. 

Let us now utilize what we have said about our oscillators in order to find out what such functional is. First of all, we imagine that we have a torus, 
\beq x^1 +L_1 = x^1 \; , \; x^2 + L_2 = x^2 \; , \; x^3 + L_3 = x^3 \eeq
and then we define the momentum $\vec{p}_{abc}$ as 
\beq \vec{p}_{abc} = \bigg( \frac{2 \pi a}{L^1}, \frac{2 \pi b}{L^2}, \frac{2 \pi c}{L^3} \bigg) \eeq
Furthermore, in order to simplify notation, we will assume some sort of sequence 
\beq \{\cdots, (a_{-2}, b_{-2}, c_{-2}), (a_{-1}, b_{-1}, c_{-1}), (a_0,b_0,c_0), (a_1, b_1, c_1), (a_2, b_2, c_2), \cdots \} \eeq
such that the following conditions hold:
\beq (a_0, b_0, c_0) = (0,0,0) \eeq
\beq (a_{-k}, b_{-k}, c_{-k}) = (-a_k, -b_k, -c_k) \eeq 
\beq \forall (d,e,f) \neq (0,0,0) \; [\exists k ((a_k,b_k,c_k) = (d,e,f))] \eeq
\beq \forall k \neq l ((a_k,b_k,c_k) \neq (a_l,b_l,c_l)) \eeq
Once we have done it, we will define $\vec{p}_k$ as 
\beq \vec{p}_k = \vec{p}_{a_k,b_k,c_k} \eeq
Thus, in particular,
\beq \vec{p}_0 = \vec{p}_{000} = \vec{0} \eeq
Any given $\phi (x)$ can be represented as 
\beq \phi (\vec{x}) = \sqrt{\frac{2}{L_1L_2L_3}} \bigg( \frac{R_0 (\phi)}{2} + \sum R_k (\phi) \cos (\vec{p}_k \cdot \vec{x} - \Theta_k (\phi)) \bigg) \eeq
where $R_0 (\phi)$, $R_k (\phi)$ and $\Theta_k (\phi)$ are given by 
\beq R_0 (\phi) = R_{000} (\phi) = R_{\vec{0}} (\phi) = R_{\vec{p}_0} (\phi) = \frac{1}{\sqrt{L_1L_2L_3}} \bigg\vert \int d^3 x \; \phi (\vec{x})   \bigg\vert \label{R000} \eeq
\beq k \neq 0 \Longrightarrow R_k (\phi) = R_{a_kb_kc_k} (\phi)= R_{\vec{p}_k} (\phi) = \sqrt{\frac{2}{L_1L_2L_3}} \bigg\vert \int d^3 x \; \phi (\vec{x}) e^{i \vec{p}_k \cdot \vec{x}} \bigg\vert \label{RABC} \eeq
\beq k \neq 0 \Longrightarrow \Theta_k (\phi) = \Theta_{a_kb_kc_k} (\phi)= \Theta_{\vec{p}_k} (\phi) = \Im \ln \int \phi (\vec{x}) e^{i \vec{p}_k \cdot \vec{x}} d^3 x \label{ThetaABC} \eeq
This implies that 
\beq R_k (\phi) = R_{-k} (\phi) \; , \; \Theta_k (\phi) = - \Theta_{-k} (\phi) \eeq
We have used the letters $R$ and $\Theta$ for a reason. The above can be interpreted as a single 1D oscillator, corresponding to zero momentum, and infinitely many 2D oscillators, corresponding to all of the allowed non-zero momenta. The 2D oscillator number $k$ simultaneously describes all particles with momentum $\vec{p}_k$ as well as all particles with momentum $- \vec{p}_k$. Therefore, the 2D oscillator number $k$ and 2D oscillator number $-k$ is the very same thing. On the other hand, 1D oscillator is assigned number $0$ (although it doesn't have to since we only have one 1D oscillator anyway) and it describes particles with zero momentum. By noticing the difference in coefficient of $\sqrt{2}$ between Eq \ref{R000} and \ref{RABC} among other similar differences, one can see that the "mass" of 1D oscillator is different from the "masses" of 2D ones:
\beq \mu_0 = \frac{1}{2} \; , \; \mu_k = 1 \; , \; k \neq 0 \eeq
These are not to be confused with the mass of the particle which is \emph{not} equal to either of those (indeed, the particle mass has a dimension, while the above so-called masses are dimensionless). In particular, the "mass" of the particle becomes the "frequency" of the oscillator, while the "mass" of the oscillator remains either $1$ or $1/2$ as described. In order not to confuse the two, we will denote the mass of the particle by $m$ and the mass of the oscillator by $\mu$. 

As we mentioned earlier, the fact that in quantum mechanics the oscillator states can be represented as functions implies that in quantum field theory they can be represented as functionals where $x$ is being replaced by $\phi$. In case of any given 2D oscillator, we replace the polar coordinates $(r, \theta)$ used in previous section with $(R_k (\phi), \Theta_k (\phi))$. On the other hand, for 1D oscillator we replace $x$ used in previous section with $R_0 (\phi)$. But, \emph{in contrast to} previous section, we will take infinite product of infinitely many oscillators. In particular, the functional of the vacuum state is the product of the wavefunction of 1D vacuum corresponding to a statement "there are no particles with momentum $\vec{0}$" with infinitely many wavefunctions of 2D vacua, corresponding to the statement "there are no particles with momentum $\vec{p}_k$" for any given $k$. Thus, the functional for vacuum state is given by 
\beq \psi_{\vert \Omega \rangle} (\phi)  = \bigg( \frac{m}{2 \pi} \bigg)^{1/4} e^{-m R_0^2 (\phi) /4}   \prod_{k \geq 1} \Bigg( \frac{(m^2+ \vert \vec{p}_k \vert^2)^{1/4}}{\pi^{1/2}}    e^{- \sqrt{m^2+ \vert \vec{p}_k \vert^2}  R_0^2 (\phi)/2} \Bigg)   \eeq
The reason we have taken a product over $k \geq 1$ instead of $k \neq 0$ is because of the remark that we have made earlier that an oscillator number $k$ is the same as an oscillator number $-k$, so we don't want to count the same oscillator twice. In other words, we could have either taken a product over $k \geq 1$ or over $k \leq -1$, but not both. The answer in case of either choice would be identical. Anyway, after substituting the equation for $R_0 (\phi)$ this becomes 
\beq \psi_{\vert \Omega \rangle} (\phi)  = \bigg( \frac{m}{2 \pi} \bigg)^{1/4} \exp \bigg( -\frac{m}{4L_1L_2L_3} \bigg\vert \int d^3 x_0 \; \phi (\vec{x}_0) \bigg\vert^2 \bigg) \times  \eeq
\beq \times   \prod_{k \geq 1} \Bigg( \frac{(m^2+ \vert \vec{p}_k \vert^2)^{1/4}}{\pi^{1/2}}    \exp \bigg( -  \frac{\sqrt{m^2+ \vert \vec{p}_k \vert^2}}{L_1L_2L_3}  \bigg\vert \int d^3 x_{ABC} \; \phi (\vec{x}_k) e^{i \vec{p}_{ABC} \cdot \vec{x}_k} \bigg\vert ^2  \bigg) \Bigg) \nonumber  \eeq

Now, when we are looking at excited states, we have to distinguish $\vec{p} = \vec{0}$ from $\vec{p} \neq \vec{0}$ as well as $\vec{p}_i = - \vec{p_j}$ from $\vec{p}_i \neq - \vec{p}_j$ (or, equivalently, $i=-j$ from $i \neq -j$). The reason is that $\vec{p}= \vec{0}$ forms 1D oscillator, while $\{\vec{p}_k, - \vec{p}_k \} = \{ \vec{p}_k, \vec{p}_{-k} \}$ forms 2D oscillator for any given $k \neq 0$.  More precisely, in all cases we have a product of a single 1D oscillator with infinitely many 2D ones. But the question is which ones are kept in a ground state and which ones are raised to excited states. The particle with zero momentum raises 1D oscillator to first excited state while leaving all of the 2D oscillators in a ground state, while the particle with non-zero momentum raises one of the 2D oscillators into first excited state, while keeping both the 1D oscillator, as well as all other 2D oscillators (except for the aforementioned one) in the ground state. 

Let us now show exactly how it works. Suppose we have one particle with zero momentum. Since all of the 2D oscillators are left in a ground state, the product of their functionals can be absorbed into $\psi_{\vert \Omega \rangle} (\phi)$. On the other hand, 1D oscillator is raised to first excited state. \emph{But} the comparison of Eq \ref{1D1} to Eq \ref{1D0} tell us that Eq \ref{1D1} has the same Gaussian as Eq \ref{1D0} does, times an extra factor. Thus, the Gaussian part from the Eq \ref{1D1} can, similarly, be absorbed into $\psi_{\vert \Omega \rangle} (\phi)$, and the extra factor is the only thing we are left with. Thus, we write down the functional to be 
\beq \psi_{\vert p=0 \rangle} (\phi) = \sqrt{m} R_0 (\phi) \psi_{\vert \Omega \rangle} (\phi) \label{Functional10} \eeq
where we have obtained the coefficient of $\sqrt{m}$ from 
\beq \sqrt{2 \mu_0 \omega_0} = \sqrt{2 \cdot \frac{1}{2} \cdot \omega_0} = \sqrt{\omega_{000}} = \sqrt{m} \label{Coeff1}\eeq 
Now, if we consider non-zero momentum, then the 1D oscillator is left in ground state, thus it is \emph{fully} absorbed into $\psi_{\vert 0 \rangle} (\phi)$, but \emph{one of} the 2D oscillators is now in a first excited state and is no longer fully absorbed the way it was previously. The comparison of Eq \ref{2D1-1} and Eq \ref{2D0} tells us that the Gaussian part of said 2D oscillator can still be absorbed into $\psi_{\vert 0 \rangle} (\phi)$, but then there is an extra coefficient that can't be. So, as before, take an extra coefficient \emph{without} Gaussian; but, this time, said extra coefficient is coming from 2D oscillator rather than 1D. Another thing that is important to stress is that, even though we have infinite product of 2D oscillators, we do \emph{not} have a product of "extra coefficients". The reason is that $\infty-1$ of those 2D oscillators are still in a ground state, and it is only \emph{one} 2D oscillator that has been raised to the first excited state. Thus $\psi_{\vert 0 \rangle} (\phi)$ \emph{fully} absorbs $\infty-1$ of $2D$ oscillators and "partially" absorbs the remaining one, so we have to include only one extra coefficient. Thus, we have 
\beq k \neq 0 \Longleftrightarrow \vec{p}_k \neq \vec{0} \Longrightarrow \psi_{\vert -p_k \rangle } (\phi) = (m^2 + \vert \vec{p}_k \vert^2)^{1/4}  R_k (\phi)  e^{-i \Theta_k (\phi)} \psi_{\vert \Omega \rangle} (\phi) \label{Functional1-1} \eeq
where we have obtained the coefficient $ (m^2 + \vert \vec{p}_k \vert^2)^{1/4}$ via
\beq  k \neq 0 \Longleftrightarrow \vec{p}_k \neq \vec{0} \Longrightarrow \sqrt{\mu \omega_k} = \sqrt{1 \cdot \omega_k} = \sqrt{\omega_k} =  (m^2 + \vert \vec{p}_k \vert^2)^{1/4} \label{Coeff2} \eeq
In other words, the coefficient happens to be the same as previously, but for different reasons: on the one hand, instead of $\sqrt{2 \mu \omega}$ we now have $\sqrt{\mu \omega}$ and, on the other hand, instead of $\mu = 1/2$ we now have $\mu = 1$. In retrospect, this is not an accident, since, in Cartesian coordinates, 2D oscillator is simply a product of two 1D ones. Finally, identical argument in which, instead of comparing Eq \ref{2D1-1} to Eq \ref{2D0} we compare Eq \ref{2D11} to Eq \ref{2D0}, tells us that 
\beq k \neq 0 \Longleftrightarrow \vec{p}_k \neq \vec{0} \Longrightarrow \psi_{\vert p_k \rangle } (\phi) = (m^2 + \vert \vec{p}_k \vert^2)^{1/4}   R_k (\phi)  e^{i \Theta_k (\phi)} \psi_{\vert \Omega \rangle} (\phi) \label{Functional11} \eeq
The fact that 
\beq k \neq 0  \Longleftrightarrow \vec{p}_k \neq \vec{0} \Longrightarrow \Theta_k (\phi) = - \Theta_k (\phi) \label{MinusTheta} \eeq
allows us to combine Eq \ref{Functional1-1} and Eq \ref{Functional11} into a single equation. Furthermore, comparison this equation to Eq \ref{Functional10} allows us to combine all three of them into a single equation, which would be the same as Eq \ref{Functional11} with $k \neq 0$ condition being dropped: 
\beq \forall k \; \Big( \psi_{\vert p_{abc} \rangle } (\phi) = \sqrt{m}  R_k (\phi)  e^{i \Theta_k (\phi)} \psi_{\vert \Omega \rangle} (\phi) \Big) \label{forall} \eeq
However, due to the fact that the equation for $R_0$ and $R_k$ differ by $\sqrt{2}$, if we are going to explicitly plug in the expressions for the latter, we would likewise have $\sqrt{2}$ difference in overall coefficient (apart from the fact that in the zero case we skip $e^{i \Theta (\phi)}$ seeing that it is equal to $1$). Thus, in case of zero momentum we have 
\beq \psi_{\vert p =0 \rangle} (\phi) =  \bigg( \sqrt{\frac{m}{L_1L_2L_3}} \bigg\vert \int d^3 x' \phi (\vec{x}')  \bigg\vert \bigg) \bigg( \bigg( \frac{m}{2 \pi} \bigg)^{1/4} \exp \bigg( -\frac{m}{4 L_1 L_2 L_3} \bigg\vert \int d^3 x_0 \; \phi (\vec{x}_0) \bigg\vert ^2 \bigg) \bigg) \times  \nonumber \eeq
\beq \times   \prod_{k \geq 1} \Bigg( \frac{(m^2+ \vert \vec{p}_k \vert^2)^{1/4}}{\pi^{1/2}}    \exp \bigg( -  \frac{\sqrt{m^2+ \vert \vec{p}_k \vert^2}}{L_1L_2L_3}  \bigg\vert \int d^3 x_k \; \phi (\vec{x}_k) e^{i \vec{p}_k \cdot \vec{x}_k} \bigg\vert ^2  \bigg) \Bigg)  \eeq
while in case of nonzero momentum we obtain 
\beq k \neq 0 \Longleftrightarrow \vec{p}_k \neq \vec{0} \Longrightarrow \nonumber \eeq
\beq \Longleftrightarrow \psi_{\vert p_k \rangle} (\phi) =  \sqrt{\frac{2 }{L_1L_2L_3}} (m^2+ \vert \vec{p}_k \vert^2 )^{1/4} \bigg\vert \int d^3 x' \phi (\vec{x}') e^{i \vec{p}_k \cdot \vec{x}'} \times \nonumber \eeq
\beq \times \bigg\vert  \exp \bigg(\ii   \Im \ln \int d^3 x'' \; \phi (\vec{x}) e^{i \vec{p}_k \cdot \vec{x}''} \bigg) \times \nonumber \eeq
\beq \times \bigg( \frac{m}{2 \pi} \bigg)^{1/4} \exp \bigg( -\frac{m}{4 L_1 L_2 L_3} \bigg\vert \int d^3 x_0 \; \phi (\vec{x}_0) \bigg\vert ^2 \bigg) \times  \eeq
\beq \times   \prod_{k \geq 1} \Bigg( \frac{(m^2+ \vert \vec{p}_k \vert^2)^{1/4}}{\pi^{1/2}}    \exp \bigg( -  \frac{\sqrt{m^2+ \vert \vec{p}_k \vert^2}}{L_1L_2L_3}  \bigg\vert \int d^3 x_k \; \phi (\vec{x}_k) e^{i \vec{p}_k \cdot \vec{x}_k} \bigg\vert ^2  \bigg) \Bigg) \nonumber  \eeq
Let us now move to two particle case. If we have two particles of zero momentum, we have to raise 1D oscillator to second excited state while keeping all of the 2D oscillators in a ground state. Thus, all of the 2D oscillators are absorbed in $\psi_{\vert \Omega \rangle} (\phi)$ while 1D oscillator, via a comparison of Eq \ref{1D2} to Eq \ref{1D0}, gives us
\beq \psi_{\vert 00 \rangle} (\phi) = \frac{m R^2_0 (\phi)-1}{\sqrt{2}} \psi_{\vert \Omega \rangle} (\phi) \eeq
which, upon substitution of $R (\phi)$ as well as $\psi_{\Omega} (\phi)$ becomes
\beq \psi_{\vert 00 \rangle} (\phi) = \frac{1}{\sqrt{2}} \bigg( \frac{m}{L_1L_2L_2} \bigg\vert \int d^3 x'  \; \phi^2 (\vec{x}') \bigg\vert^2 -1 \bigg) \times \nonumber \eeq
\beq \times \bigg( \frac{m}{2 \pi} \bigg)^{1/4} \exp \bigg( -\frac{m}{4L_1L_2L_3} \bigg\vert \int d^3 x_0 \; \phi (\vec{x}_0) \bigg\vert^2 \bigg) \times  \eeq
\beq \times   \prod_{k \geq 1} \Bigg( \frac{(m^2+ \vert \vec{p}_k \vert^2)^{1/4}}{\pi^{1/2}}    \exp \bigg( -  \frac{\sqrt{m^2+ \vert \vec{p}_k \vert^2}}{L_1L_2L_3}  \bigg\vert \int d^3 x_k \; \phi (\vec{x}_k) e^{i \vec{p}_k \cdot \vec{x}_k} \bigg\vert ^2  \bigg) \Bigg) \nonumber  \eeq
In case of $\vec{p}_k \neq \vec{0}$ and $\vec{p}_l \neq \vec{0}$ (or, equivalently, $k \neq 0$ and $l \neq 0$), we have to use 2D oscillator. If $\vec{p}_k = \pm \vec{p}_l$ (or, equivalently, $k = \pm l$), then we have second excited state of the 2D oscillator number $k$ (which coincides with 2D oscillator number $-k$) and ground state of all the other ones; thus, we use Eq \ref{2D20}, \ref{2D22} and \ref{2D2-2}. On the other hand, if $\vec{p}_k \neq \pm \vec{p}_l$ (or, equivalently, $k \neq \pm l$) then we have two of the 2D oscillators raised to the first excited state (namely, 2D oscillators number $k$ and $l$ which coincide with oscillators number $-k$ and $-l$, respectively), and everything else kept on a ground state; thus, we use Eq \ref{2D11} and \ref{2D1-1}. And, finally, if we have $\vec{p}_k = \vec{0}$ and $\vec{p}_l \neq \vec{0}$ (or, equivalently, $k=0$ and $l \neq 0$), then 1D oscillator (which is always number $0$ by default since there is only one 1D oscillator available altogether), as well as the 2D oscillators number $l$ (which coincides with 2D oscillator number $-l$), will be in the first excited state, and all other 2D oscillators in ground state; thus, we combine Eq \ref{1D1} with either \ref{2D11} or \ref{2D1-1}. Going back to $\vec{p}_k = \pm \vec{p}_l$ (or, equivalently, $k = \pm l$), we have to distinguish the case of $\vec{p}_k =  \vec{p}_l$ (or, equivalently, $k=l$) from $\vec{p}_k = - \vec{p}_l$ (or, equivalently, $k=-l$). In the case of $\vec{p}_k =  \vec{p}_l$ (or, equivalently, $k=l$), the total linear momentum is $2 \vec{p}_k$, corresponding to angular momentum $\pm 2$ (where $\pm$ becomes $+$ if $k>0$ and $-$ if $k<0$) thus we have to use either Eq \ref{2D22} or \ref{2D2-2}. On the other hand, in the case of $\vec{p}_k = - \vec{p}_l$ (or, equivalently, $k<l$), we have total linear momentum zero, corresponding to zero angular momentum. Thus, we have to use Eq \ref{2D20}. By keeping in mind everything we said so far, we obtain the following functionals:
\beq k \neq 0 \Longleftrightarrow \vec{p}_k \neq \vec{0} \Longrightarrow \psi_{\vert p_k, -p_k \rangle} (\phi) = \Big( \sqrt{\vert \vec{p}_k \vert^2 +m^2} R^2_k (\phi) - 1 \Big) \psi_{\vert \Omega \rangle} (\phi) \label{1} \eeq 
\beq k \neq 0 \Longleftrightarrow \vec{p}_k \neq \vec{0} \Longrightarrow \psi_{\vert p_k, p_k \rangle} (\phi) = \sqrt{\frac{\vert \vec{p}_k \vert^2 +m^2}{2}} R_k^2 (\phi) e^{2i \Theta_k (\phi)} \psi_{\vert \Omega \rangle} (\phi) \label{2} \eeq
\beq k \neq \pm l \Longleftrightarrow \vec{p}_k \neq \pm \vec{p}_l \Longrightarrow \nonumber \eeq
\beq \Longrightarrow \psi_{\vert p_k p_l \rangle} (\phi) = (m^2 + \vert \vec{p}_k \vert^2)^{1/4} (m^2 + \vert \vec{p}_l \vert^2)^{1/4} R_k (\phi) R_l (\phi) e^{i \Theta_k (\phi)} e^{i \Theta_l (\phi)} \psi_{\vert \Omega \rangle} (\phi) \label{3abc} \eeq
The way we avoided much longer list is that we have used the kind of argument that allowed us to combine Eq \ref{Functional10}, \ref{Functional1-1} and \ref{Functional11} into a single equation, \ref{forall}. In particular, we utilized Eq \ref{MinusTheta} as well as the similarity between Eq \ref{Coeff1} and \ref{Coeff2} .Clearly, it we still have to distinguish some cases, but at least we can shorten the list of cases to be compared. Now, plugging in $R (\phi)$ and $\psi_{\vert \Omega \rangle} (\phi)$ into Eq \ref{1} and \ref{2} is straightforward since, in both cases, we have to use the expression for $R$ given for non-zero momentum. Thus, Eq \ref{1} becomes 
\beq k \neq 0 \Longleftrightarrow \vec{p}_k \neq \vec{0} \Longrightarrow \nonumber \eeq
\beq \Longrightarrow \psi_{\vert p_k, -p_k \rangle} (\phi) =  \bigg( \frac{2 \sqrt{\vert \vec{p}_k \vert^2+m^2}}{L_1L_2L_2} \bigg\vert \int d^3 x'  \; \phi^2 (\vec{x}') e^{i \vec{p}_k \cdot \vec{x}'} \bigg\vert^2 -1 \bigg) \times \nonumber \eeq
\beq \times  \bigg( \frac{m}{2 \pi} \bigg)^{1/4} \exp \bigg( -\frac{m}{4L_1L_2L_3} \bigg\vert \int d^3 x_0 \; \phi (\vec{x}_0) \bigg\vert^2 \bigg) \times  \nonumber \eeq
\beq \times   \prod_{j\geq 0} \Bigg( \frac{(m^2+ \vert \vec{p}_j \vert^2)^{1/4}}{\pi^{1/2}}    \exp \bigg( -  \frac{\sqrt{m^2+ \vert \vec{p}_j \vert^2}}{L_1L_2L_3}  \bigg\vert \int d^3 x_j \; \phi (\vec{x}_j) e^{i \vec{p}_j \cdot \vec{x}_k} \bigg\vert ^2  \bigg) \Bigg)   \eeq
while Eq \ref{2} becomes 
\beq k \neq 0 \Longleftrightarrow \vec{p}_k \neq 0 \Longrightarrow \nonumber \eeq
\beq \Longrightarrow \psi_{\vert p_k, p_k \rangle} (\phi) =  \bigg( \frac{ \sqrt{2 (\vert \vec{p}_k \vert^2+m_k^2)}}{L_1L_2L_2} \bigg\vert \int d^3 x'  \; \phi^2 (\vec{x}') e^{i \vec{p}_k \cdot \vec{x}'}\bigg\vert^2  \bigg) \times \nonumber \eeq
\beq \times  \exp \bigg(2 \ii  \Im \ln \int d^3 x'' \; \phi (\vec{x}'') e^{i \vec{p}_k \cdot \vec{x}''} \bigg) \times \nonumber \eeq
\beq \times \bigg( \frac{m}{2 \pi} \bigg)^{1/4} \exp \bigg( -\frac{m}{4L_1L_2L_3} \bigg\vert \int d^3 x_0 \; \phi (\vec{x}_0) \bigg\vert^2 \bigg) \times  \eeq
\beq \times   \prod_{j \geq 1} \Bigg( \frac{(m^2+ \vert \vec{p}_j \vert^2)^{1/4}}{\pi^{1/2}}    \exp \bigg( -  \frac{\sqrt{m^2+ \vert \vec{p}_j \vert^2}}{L_1L_2L_3}  \bigg\vert \int d^3 x_j \; \phi (\vec{x}_j) e^{i \vec{p}_j \cdot \vec{x}_j} \bigg\vert ^2  \bigg) \Bigg) \nonumber  \eeq
On the other hand, Eq \ref{3abc} requires some extra care since it is used both for the case where both momenta are non-zero as well as the case where one of them is zero and the other is non-zero (the case where both are zero is ruled out since we have stated that the two momenta are not equal to each other). The situation where neither of the two momenta is zero is described as 
\beq \vec{0} \neq \vec{p}_k \neq \pm \vec{p}_l \neq \vec{0}  \Longleftrightarrow 0 \neq k \neq \pm l \neq 0 \Longrightarrow \nonumber \eeq
\beq \Longrightarrow \psi_{\vert p_k p_l \rangle} (\phi) = \bigg( \sqrt{\frac{2 }{L_1L_2L_3}} (m^2+ \vert \vec{p}_k \vert^2 )^{1/4}  \bigg\vert \int d^3 x' \; \phi (\vec{x}') e^{i\vec{p}_k \cdot \vec{x}'} \bigg\vert \bigg) \times \nonumber \eeq
\beq \times \bigg( \sqrt{\frac{2 }{L_1L_2L_3}} (m^2+ \vert \vec{p}_l \vert^2 )^{1/4}  \bigg\vert \int d^3 x'' \; \phi (\vec{x}'') e^{i\vec{p}_l \cdot \vec{x}''} \bigg\vert \bigg) \times \nonumber \eeq
\beq  \times \bigg[ \exp \bigg( \ii  \Im \ln \int d^3 x''' \; \phi (\vec{x}''') e^{i \vec{p}_k \cdot \vec{x}'''}  \bigg) \bigg]\bigg[ \exp \bigg( \ii  \Im \ln \int d^3 x'''' \phi (\vec{x}) e^{i \vec{p}_l \cdot \vec{x}''''} \bigg) \bigg] \times \eeq 
\beq \times \bigg( \frac{m}{2 \pi} \bigg)^{1/4} \exp \bigg( -\frac{m}{4L_1L_2L_3} \bigg\vert \int d^3 x_0 \; \phi (\vec{x}_0) \bigg\vert^2 \bigg) \times  \nonumber \eeq
\beq \times   \prod_{j \geq 1} \Bigg( \frac{(m^2+ \vert \vec{p}_j \vert^2)^{1/4}}{\pi^{1/2}}    \exp \bigg( -  \frac{\sqrt{m^2+ \vert \vec{p}_j \vert^2}}{L_1L_2L_3}  \bigg\vert \int d^3 x_j \; \phi (\vec{x}_j) e^{i \vec{p}_j \cdot \vec{x}_j} \bigg\vert ^2  \bigg) \Bigg) \nonumber  \eeq
On the other hand, the situation where one of the momenta is zero is described as 
\beq l \neq 0 \Longleftrightarrow \vec{p}_l \neq \vec{0} \Longrightarrow \nonumber \eeq
\beq \Longrightarrow \psi_{\vert 0p_l  \rangle} (\phi) = \bigg( \sqrt{\frac{m}{L_1L_2L_3}} \bigg\vert \int d^3 x' \; \phi (\vec{x}') \bigg\vert \bigg) \times \nonumber \eeq
\beq \times \bigg( \sqrt{\frac{2 }{L_1L_2L_3}} (m^2+ \vert \vec{p}_l \vert^2 )^{1/4}  \bigg\vert \int d^3 x'' \; \phi (\vec{x}'') e^{i\vec{p}_l \cdot \vec{x}''} \bigg\vert \bigg) \times \nonumber \eeq
\beq  \times \bigg[ \exp \bigg( \ii  \Im \ln \int d^3 x''' \; \phi (\vec{x}) e^{i \vec{p}_l \cdot \vec{x}'''}  \bigg) \bigg] \times \eeq 
\beq \times \bigg( \frac{m}{2 \pi} \bigg)^{1/4} \exp \bigg( -\frac{m}{4L_1L_2L_3} \bigg\vert \int d^3 x_0 \; \phi (\vec{x}_0) \bigg\vert^2 \bigg) \times  \nonumber \eeq
\beq \times   \prod_{j \geq 1} \Bigg( \frac{(m^2+ \vert \vec{p}_j \vert^2)^{1/4}}{\pi^{1/2}}    \exp \bigg( -  \frac{\sqrt{m^2+ \vert \vec{p}_j \vert^2}}{L_1L_2L_3}  \bigg\vert \int d^3 x_j \; \phi (\vec{x}_j) e^{i \vec{p}_j \cdot \vec{x}_j} \bigg\vert ^2  \bigg) \Bigg) \nonumber  \eeq
This procedure can be extended to general particle numbers by utilizing Eq \ref{nthState1D} and \ref{nLState2D}. In light of the fact that particles are not distinguishable, combined with the fact that we have aforegiven list of allowed momenta, in order to specify a state we simply have to list the particle numbers corresponding to each allowed momentum. We will denote the number of particles with momentum $k$ by $\sharp (\vec{p}_k)$. Since zero momentum corresponds to 1D oscillator and non-zero momentum corresponds to 2D, we use Eq \ref{nthState1D} to account for arbitrary number of particles with zero momentum and Eq \ref{nLState2D} to account for the arbitrary number of particles of non-zero momentum. Since there is only one zero momentum state and infinitely many non-zero ones, we take a product of one copy of Eq \ref{nthState1D} with arbitrary many copies of Eq \ref{nLState2D}, each copy being "adjusted" for different momentum. Thus, we obtain
\beq \psi_{\vert \sharp (\vec{0}) = n_0, \sharp (\vec{p}_1) = n_1, \sharp (-\vec{p}_1) = n_{-1}, \sharp (\vec{p}_2) = n_2, \sharp (- \vec{p}_2) = n_{-2}, \cdots \rangle}    (\phi) =  \nonumber \eeq
\beq = \bigg( \sqrt{n_0!} \bigg( \frac{m}{2 \pi} \bigg)^{1/4} e^{-m R_0^2 (\phi) /4} \sum_{C_0=0}^{\lfloor n/2 \rfloor} \frac{(-1)^{C_0} m^{\frac{n_0}{2} - C_0}}{2^{C_0} C_0! (n_0-2C_0)!} R_0^{n_0-2C_0} (\phi) \bigg) \times \label{GenFunctional1} \eeq
\beq \times \prod_{k} \Bigg( \frac{(m^2+ \vert \vec{p}_k \vert^2)^{1/4}}{\pi^{1/2}}  \sqrt{2^{n_k+n_{-k}} n_k! n_{-k}!}  e^{- \sqrt{m^2 + \vert \vec{p}_k \vert^2} R_0^2 (\phi)/2}   \times \nonumber \eeq
\beq \times \sum_{C_k=0}^{\min (n_k, n_{-k})} \frac{(-1)^{C_k} 2^{\frac{n_k+n_{-k}}{2}-C_k} (m^2 + \vert \vec{p}_k \vert^2)^{\frac{n_k+n_{-k}}{4} - \frac{C_k}{2}} R_k^{n-2C_k} (\phi) e^{i (n_k-n_{-k}) \Theta_k (\phi)}}{2^{n_k+n_{-k}-C_k} C! (n_k-C_k)! (n_{-k} -C_k )!}  \Bigg)  \nonumber \eeq
Now, if we plug in zero particle numbers, we will obtain the functional for vacuum state: 
\beq \psi_{\vert \Omega \rangle} (\phi)  = \bigg( \frac{m}{2 \pi} \bigg)^{1/4} e^{-m R_0^2 (\phi) /4}   \prod_{k} \Bigg( \frac{(m^2+ \vert \vec{p}_k \vert^2)^{1/4}}{\pi^{1/2}}    e^{- \sqrt{m^2+ \vert \vec{p}_k \vert^2}  R_0^2 (\phi)/2} \Bigg)  \label{VacFunctional} \eeq
and, therefore, by absorbing some of Eq \ref{GenFunctional1} into $\psi_{\vert \Omega \rangle} (\phi)$ via Eq \ref{VacFunctional}, the general functional can be rewritten as 
\beq  \psi_{\vert \sharp (\vec{0}) = n_0, \sharp (\vec{p}_1) = n_1, \sharp (-\vec{p}_1) = n_{-1}, \sharp (\vec{p}_2) = n_2, \sharp (- \vec{p}_2) = n_{-2}, \cdots \rangle} (\phi) =  \nonumber \eeq
\beq = \psi_{\vert \Omega \rangle} (\phi) \bigg( \sqrt{n_0!} \sum_{C_0=0}^{\lfloor n/2 \rfloor} \frac{(-1)^{C_0} m^{\frac{n_0}{2} - C_0}}{2^{C_0} C_0! (n_0-2C_0)!} R_0^{n_0-2C_0} (\phi) \bigg) \times \eeq
\beq \times \prod_{k} \Bigg(   \sqrt{2^{n_k+n_{-k}} n_k! n_{-k}!}     \sum_{C_k=0}^{\min (n_k, n_{-k})} \frac{(-1)^{C_k} 2^{\frac{n_k+n_{-k}}{2}-C_k} (m^2 + \vert \vec{p}_k \vert^2)^{\frac{n_k+n_{-k}}{4} - \frac{C_k}{2}} R_k^{n-2C_k} (\phi) e^{i (n_k-n_{-k}) \Theta_{p_k} (\phi)}}{2^{n_k+n_{-k}-C_k} C! (n_k-C_k)! (n_{-k} -C_k )!}  \Bigg)  \nonumber \eeq
and if we plug in the equations for $\psi_{\vert \Omega \rangle} (\phi)$ as well as $R_k (\phi)$ we obtain 
\beq  \psi_{\vert \sharp (\vec{0}) = n_0, \sharp (\vec{p}_1) = n_1, \sharp (-\vec{p}_1) = n_{-1}, \sharp (\vec{p}_2) = n_2, \sharp (- \vec{p}_2) = n_{-2}, \cdots \rangle}    (\phi)  \nonumber \eeq
\beq = \Bigg[ \bigg( \frac{m}{2 \pi} \bigg)^{1/4} \exp \bigg( -\frac{m}{4L_1L_2L_3} \bigg\vert \int d^3 x_0 \; \phi (\vec{x}_0) \bigg\vert^2 \bigg) \times  \nonumber \eeq
\beq \times   \prod_{k \geq 1} \Bigg( \frac{(m^2+ \vert \vec{p}_k \vert^2)^{1/4}}{\pi^{1/2}}    \exp \bigg( -  \frac{\sqrt{m^2+ \vert \vec{p}_k \vert^2}}{L_1L_2L_3}  \bigg\vert \int d^3 x_k \; \phi (\vec{x}_k) e^{i \vec{p}_k \cdot \vec{x}_k} \bigg\vert ^2  \bigg) \Bigg) \Bigg] \times \nonumber  \eeq
\beq \times \bigg( \sqrt{n_0!} \sum_{C_0=0}^{\lfloor n/2 \rfloor} \frac{(-1)^{C_0} m^{\frac{n_0}{2} - C_0}}{2^{C_0} C_0! (n_0-2C_0)!} \bigg\vert \int d^3 x' \; \phi (\vec{x}') \bigg\vert^{n_0-2C_0} \bigg) \times \nonumber \eeq
\beq \times \prod_{k} \Bigg(   \sqrt{2^{n_k+n_{-k}} n_k! n_{-k}!}     \sum_{C_k=0}^{\min (n_k, n_{-k})} \frac{(-1)^{C_k} 2^{\frac{n_k+n_{-k}}{2}-C_k} (m^2 + \vert \vec{p}_k \vert^2)^{\frac{n_k+n_{-k}}{4} - \frac{C_k}{2}} }{2^{n_k+n_{-k}-C_k} C! (n_k-C_k)! (n_{-k} -C_k )!}  \times   \nonumber \eeq
\beq \times \bigg\vert \int d^3 x'' \; \phi (\vec{x}'') e^{i \vec{p}_k \cdot \vec{x}''} \bigg\vert^{n_k-2C_k} \exp \bigg(i (n_k-n_{-k}) \Im \ln \int d^3 x''' \; \phi (\vec{x}''') e^{i \vec{p}_k \cdot \vec{x}'''} \bigg) \bigg) \label{ExplicitFunctional} \eeq

Now, in order to write down creation and annihilation operators in differential form, first of all, let us define the derivatives. One can show that 
\beq \Theta_k (\phi + \epsilon \cos (\vec{k} \cdot \vec{x} - \Theta_l (\phi))) = \Theta_k (\phi)  + 0 (\epsilon^2) \eeq
\beq \Theta_k (\phi + \epsilon \sin (\vec{k} \cdot \vec{x} - \Theta_l (\phi))) = \Theta_{abc} (\phi) + \epsilon \; \frac{ \delta^k_l}{R_k (\phi)} \sqrt{\frac{L_1L_2L_3}{2}} + 0 (\epsilon^2) \eeq
\beq R_k (\phi + \epsilon \cos (\vec{k} \cdot \vec{x} - \Theta_l (\phi))) = R_k (\phi) + \epsilon \delta^k_l  \sqrt{\frac{L_1L_2L_3}{2}} + 0 (\epsilon^2) \eeq 
\beq R_k (\phi + \epsilon \sin (\vec{k} \cdot \vec{x} - \Theta_l (\phi))) = R_k (\phi)  + 0 (\epsilon^2) \eeq 
From this, we conclude that 
\beq (\partial_{\Theta_k} \psi) (\phi) = \sqrt{\frac{2}{L_1L_2L_3}}  R_k (\phi) \lim_{\epsilon \rightarrow 0} \frac{\psi \Big(\phi +  \epsilon \sin \big(\vec{k} \cdot \vec{x} - \Theta_k (\phi) \big) \Big)- \psi (\phi)}{\epsilon} \eeq
\beq (\partial_{R_k} \psi) (\phi) = \sqrt{\frac{2}{L_1L_2L_3}} \lim_{\epsilon \rightarrow 0} \frac{\psi \Big(\phi +  \epsilon \cos \big(\vec{k} \cdot \vec{x} - \Theta_k (\phi)\big)\Big) - \psi (\phi)}{\epsilon} \eeq
By substituting 
\beq R_0 (\phi) = \frac{1}{\sqrt{L_1L_2L_3}} \bigg\vert \int d^3 x \; \phi (\vec{x})   \bigg\vert \eeq
\beq k \neq 0 \Longrightarrow R_k (\phi) = \sqrt{\frac{2}{L_1L_2L_3}} \bigg\vert \int d^3 x \; \phi (\vec{x}) e^{i \vec{p}_k \cdot \vec{x}} \bigg\vert \ \eeq
\beq k \neq 0 \Longrightarrow \Theta_k (\phi) = \Im \ln \int \phi (\vec{x}) e^{i \vec{p}_k \cdot \vec{x}} d^3 x \eeq
into the right hand side we obtain

\beq k \neq 0 \Longleftrightarrow \vec{p}_k \neq \vec{0} \Longrightarrow \nonumber \eeq
\beq \Longrightarrow (\partial_{\Theta_k} \psi) (\phi) =  \frac{2}{L_1L_2L_3} \bigg\vert \int d^3 x \; \phi (\vec{x}) e^{i \vec{p}_k \cdot \vec{x}} \bigg\vert  \times \nonumber \eeq
\beq \times   \lim_{\epsilon \rightarrow 0} \frac{\psi \Big(\phi +  \epsilon \sin \big(\vec{k} \cdot \vec{x} -  \Im \ln \int \phi (\vec{x}) e^{i \vec{p}_k \cdot \vec{x}} d^3 x  \big) \Big)- \psi (\phi)}{\epsilon} \eeq
\beq k \neq 0 \Longleftrightarrow \vec{p}_k \neq \vec{0} \Longrightarrow \nonumber \eeq
\beq \Longrightarrow  (\partial_{R_k} \psi) (\phi) = \sqrt{ \frac{2}{L_1L_2L_3} }    \lim_{\epsilon \rightarrow 0} \frac{\psi \Big(\phi +  \epsilon \cos \big(\vec{k} \cdot \vec{x} -  \Im \ln \int \phi (\vec{x}) e^{i \vec{p}_k \cdot \vec{x}} d^3 x  \big) \Big)- \psi (\phi)}{\epsilon} \eeq
\beq (\partial_{R_0} \psi) (\phi) = \frac{1}{\sqrt{L_1L_2L_3}} \lim_{\epsilon \rightarrow 0} \frac{\psi (\phi + \epsilon) - \psi (\phi)}{\epsilon} \eeq
where $\phi + \epsilon$ is merely a shift by a constant: 
\beq (\phi + \epsilon) (\vec{x}) = \epsilon + \phi (\vec{x}) \eeq
By looking at the expressions for $a_{++}$ and $a_{--}$, we read off
\beq [a_{p_k}^{\dagger} (\psi)] (\phi) = \frac{e^{i \Theta_k (\phi)}}{2} \bigg(R_k (\phi) \psi (\phi) (m^2 + \vert \vec{p}_k \vert^2)^{1/4} - \nonumber \eeq
\beq - \frac{1}{ (m^2 + \vert \vec{p}_k \vert^2)^{1/4}} (\partial_{R_k} \psi)(\phi) - \frac{i}{R_k (\phi)  (m^2 + \vert \vec{p}_k \vert^2)^{1/4}} (\partial_{\Theta_k} \psi) (\phi)\bigg) \label{AnnihilationInfiniteOriginal} \eeq
\beq [a_{p_k} (\psi)] (\phi) = \frac{e^{-i \Theta_k (\phi)}}{2} \bigg(R_k (\phi) \psi (\phi) (m^2 + \vert \vec{p}_k \vert^2)^{1/4} + \nonumber \eeq
\beq + \frac{1}{ (m^2 + \vert \vec{p}_k \vert^2)^{1/4}} (\partial_{R_{abc}} \psi) (\phi) - \frac{i}{R_k (\phi)  (m^2 + \vert \vec{p}_k \vert^2)^{1/4}} (\partial_{\Theta_k} \psi) (\phi) \bigg) \label{CreationInfiniteOriginal} \eeq
and, by substituting the expressions for $R_k$, $\Theta_k$, $\partial_{R_k}$ and $\partial_{\Theta_k}$ we obtain
\beq [a_{p_k}^{\dagger} (\psi)] (\phi) = \frac{1}{2} \exp \bigg(\ii   \Im \ln \int \phi (\vec{x}) e^{i \vec{p}_k \cdot \vec{x}} d^3 x  \bigg) \times \nonumber \eeq
\beq \times \Bigg[\sqrt{\frac{2}{L_1L_2L_3}}  (m^2 + \vert \vec{p}_k \vert^2)^{1/4} \bigg\vert \int d^3 x \; \psi (\vec{x}) e^{i \vec{p}_k \cdot \vec{x}} \bigg\vert \psi (\phi) - \eeq
\beq - \frac{1}{(m^2 + \vert \vec{p}_k \vert^2)^{1/4}} \bigg( \sqrt{\frac{2}{L_1L_2L_3}} \lim_{\epsilon \rightarrow 0} \frac{\psi \Big(\phi +  \epsilon \cos \big(\vec{k} \cdot \vec{x} - \Theta_k (\phi)\big)\Big) - \psi (\phi)}{\epsilon} \bigg) - \eeq
\beq - \frac{i}{(m^2 + \vert \vec{p}_k \vert^2)^{1/4} \sqrt{\frac{2}{L_1L_2L_3}} \bigg\vert \int d^3 x \; \psi (\vec{x}) e^{i \vec{p}_k \cdot \vec{x}} \bigg\vert} \times \nonumber \eeq
\beq \times \bigg( \frac{2}{L_1L_2L_3} \bigg\vert \int d^3 x \; \psi (\vec{x}) e^{i \vec{p}_k \cdot \vec{x}} \bigg\vert  \times \nonumber \eeq
\beq \times   \lim_{\epsilon \rightarrow 0} \frac{\psi \Big(\phi +  \epsilon \sin \big(\vec{k} \cdot \vec{x} -  \Im \ln \int \phi (\vec{x}) e^{i \vec{p}_k \cdot \vec{x}} d^3 x  \big) \Big)- \psi (\phi)}{\epsilon} \bigg) \Bigg]\eeq
The expression for the annihilation operator is the same except that the first sign is switched from minus to plus: 
\beq [a_{p_k} (\psi)] (\phi) = \frac{1}{2} \exp \bigg(\ii   \Im \ln \int \phi (\vec{x}) e^{i \vec{p}_k \cdot \vec{x}} d^3 x  \bigg) \times \nonumber \eeq
\beq \times \Bigg[\sqrt{\frac{2}{L_1L_2L_3}}  (m^2 + \vert \vec{p}_k \vert^2)^{1/4} \bigg\vert \int d^3 x \; \psi (\vec{x}) e^{i \vec{p}_k \cdot \vec{x}} \bigg\vert \psi (\phi) + \eeq
\beq + \frac{1}{(m^2 + \vert \vec{p}_k \vert^2)^{1/4}} \bigg( \sqrt{\frac{2}{L_1L_2L_3}} \lim_{\epsilon \rightarrow 0} \frac{\psi \Big(\phi +  \epsilon \cos \big(\vec{k} \cdot \vec{x} - \Theta_k (\phi)\big)\Big) - \psi (\phi)}{\epsilon} \bigg) - \eeq
\beq - \frac{i}{(m^2 + \vert \vec{p}_k \vert^2)^{1/4} \sqrt{\frac{2}{L_1L_2L_3}} \bigg\vert \int d^3 x \; \psi (\vec{x}) e^{i \vec{p}_k \cdot \vec{x}} \bigg\vert} \times \nonumber \eeq
\beq \times \bigg( \frac{2}{L_1L_2L_3} \bigg\vert \int d^3 x \; \psi (\vec{x}) e^{i \vec{p}_k \cdot \vec{x}} \bigg\vert  \times \nonumber \eeq
\beq \times   \lim_{\epsilon \rightarrow 0} \frac{\psi \Big(\phi +  \epsilon \sin \big(\vec{k} \cdot \vec{x} -  \Im \ln \int \phi (\vec{x}) e^{i \vec{p}_k \cdot \vec{x}} d^3 x  \big) \Big)- \psi (\phi)}{\epsilon} \bigg) \Bigg]\eeq
Now, if we are to look at raising and lowering operators of 1D oscillator, and use
\beq \mu_0 = \frac{1}{2} \; , \; \omega_0 = m \eeq
we will read off $a_0^{\dagger}$ and $a_0$:
\beq a_0^{\dagger} = \frac{\sqrt{m}}{2} R_0 - \frac{1}{\sqrt{m }} \partial_{R_0} \; , \; a = \frac{\sqrt{m}}{2} R_0 + \frac{1}{\sqrt{m}} \partial_{R_0} \label{CreatAnnihilZeroCont} \eeq
and, by substituting the expressions for $R_0$ and $\partial_{R_0}$ we obtain
\beq a_0^{\dagger} = \frac{\sqrt{m}}{2} \bigg( \frac{1}{\sqrt{L_1L_2L_3}} \bigg\vert \int d^3 x \; \psi (\vec{x})   \bigg\vert \bigg) - \frac{1}{\sqrt{m }}  \frac{1}{\sqrt{L_1L_2L_3}} \lim_{\epsilon \rightarrow 0} \frac{\psi (\phi + \epsilon) - \psi (\phi)}{\epsilon}  \eeq
\beq a_0^{\dagger} = \frac{\sqrt{m}}{2} \bigg( \frac{1}{\sqrt{L_1L_2L_3}} \bigg\vert \int d^3 x \; \psi (\vec{x})   \bigg\vert \bigg) +\frac{1}{\sqrt{m }}  \frac{1}{\sqrt{L_1L_2L_3}} \lim_{\epsilon \rightarrow 0} \frac{\psi (\phi + \epsilon) - \psi (\phi)}{\epsilon}  \eeq

\subsection*{6. Converting functionals into functions}

As we stated earlier, our ultimate goal is to replace $\psi (\phi)$ with $\psi (\vec{x},y)$ since the former doesn't have classical ontology while the latter does. In order to do that, we need a function $\{y \} \mapsto \{ \phi \}$. In order to introduce that function, we first postulate some \emph{fixed} field $\chi (\vec{x}, y)$ and then define $\chi_y$ as 
\beq \chi_y (\vec{x}) = \chi (\vec{x}, y) \eeq
and then define $g^{(\chi)} \colon \{ y \} \mapsto \{\phi \}$ as
\beq g^{(\chi)} (y) = \chi_y \eeq
This should enable us to replace $\psi \colon \{ \phi \} \mapsto \mathbb{C}$ to $\psi \circ g^{(\chi)}  \colon \{y \} \mapsto \mathbb{C}$ via
\beq (\psi \circ g^{(\chi)}) (y) = \psi (g^{(\chi)}) (y) = \psi (\chi_y) \eeq
This, however, is not yet what we want, since we would like to have a function of the form $\{\vec{x}, y\} \mapsto \mathbb{C}$ rather than $\{y \} \mapsto \mathbb{C}$. In order to obtain function $\{ \vec{x},y \} \mapsto \mathbb{C}$, we define
\beq \xi (\vec{x},y) = f (\vec{x}) (\psi \circ g^{(\chi)}) (y) \eeq
where $f (\vec{x})$ is a wave function corresponding to the additional particle we call a "fly". In other words, we are describing all of the particles in the universe via $\psi \circ g^{(\chi)}$ and, in addition to that, we are also describing one more particle, a fly, that can't be observed. Then the QFT state $\vert \psi (\phi) \rangle$ in conjunction with a hidden field $\chi$ and a fly with momentum $\vec{p}_{fly}$ will, indeed, be described as a function of the form $(\vec{x}, y) \mapsto \mathbb{C}$, just as we wanted: 
\beq \xi_{\chi \otimes \vert p_{fly} \rangle \otimes \vert \psi (\phi) \rangle} (\vec{x},y) = \psi (\chi_y) e^{i \vec{p}_{fly} \cdot \vec{x}} \eeq 
If we now substitute Eq \ref{ExplicitFunctional} for $\psi (\phi)$, the function over $(\vec{x},y)$ will read off as 
\beq  \xi_{\chi \otimes \vert p_{fly} \rangle \otimes\vert \sharp (\vec{0}) = n_0, \sharp (\vec{p}_1) = n_1, \sharp (-\vec{p}_1) = n_{-1}, \sharp (\vec{p}_2) = n_2, \sharp (- \vec{p}_2) = n_{-2}, \cdots \rangle}    (\vec{x},y)  \nonumber \eeq
\beq = e^{i \vec{p}_{fly} \cdot \vec{x}} \Bigg[ \bigg( \frac{m}{2 \pi} \bigg)^{1/4} \exp \bigg( -\frac{m}{4L_1L_2L_3} \bigg\vert \int d^3 x_0 \; \chi (\vec{x}_0,y) \bigg\vert^2 \bigg) \times  \nonumber \eeq
\beq \times   \prod_{k \geq 1} \Bigg( \frac{(m^2+ \vert \vec{p}_k \vert^2)^{1/4}}{\pi^{1/2}}    \exp \bigg( -  \frac{\sqrt{m^2+ \vert \vec{p}_k \vert^2}}{L_1L_2L_3}  \bigg\vert \int d^3 x_k \; \chi (\vec{x}_k,y) e^{i \vec{p}_k \cdot \vec{x}_k} \bigg\vert ^2  \bigg) \Bigg) \Bigg] \times \nonumber  \eeq
\beq \times \bigg( \sqrt{n_0!} \sum_{C_0=0}^{\lfloor n/2 \rfloor} \frac{(-1)^{C_0} m^{\frac{n_0}{2} - C_0}}{2^{C_0} C_0! (n_0-2C_0)!} \bigg\vert \int d^3 x' \; \chi (\vec{x}',y) \bigg\vert^{n_0-2C_0} \bigg) \times \nonumber \eeq
\beq \times \prod_{k} \Bigg(   \sqrt{2^{n_k+n_{-k}} n_k! n_{-k}!}     \sum_{C_k=0}^{\min (n_k, n_{-k})} \frac{(-1)^{C_k} 2^{\frac{n_k+n_{-k}}{2}-C_k} (m^2 + \vert \vec{p}_k \vert^2)^{\frac{n_k+n_{-k}}{4} - \frac{C_k}{2}} }{2^{n_k+n_{-k}-C_k} C! (n_k-C_k)! (n_{-k} -C_k )!}  \times   \nonumber \eeq
\beq \times \bigg\vert \int d^3 x'' \; \chi (\vec{x}'',y) e^{i \vec{p}_k \cdot \vec{x}''} \bigg\vert^{n_k-2C_k} \exp \bigg(i (n_k-n_{-k}) \Im \ln \int d^3 x''' \; \chi (\vec{x}''',y) e^{i \vec{p}_k \cdot \vec{x}'''} \bigg) \bigg) \label{xiGeneralFinal} \eeq
Similarly, if we want fly to be localized in space rather than momentum, we have
\beq \xi_{\chi \otimes \vert x_{fly} \rangle \otimes \vert \psi (\phi) \rangle} (\vec{x},y) = \psi (\chi_y) \delta^3 (\vec{x} - \vec{x}_{fly}) \eeq 
and then the function over $(\vec{x},y)$ will be 
\beq  \xi_{\chi \otimes \vert x_{fly} \rangle \otimes\vert \sharp (\vec{0}) = n_0, \sharp (\vec{p}_1) = n_1, \sharp (-\vec{p}_1) = n_{-1}, \sharp (\vec{p}_2) = n_2, \sharp (- \vec{p}_2) = n_{-2}, \cdots \rangle}    (\vec{x},y)  \nonumber \eeq
\beq = \delta^3 (\vec{x} - \vec{x}_{fly}) \Bigg[ \bigg( \frac{m}{2 \pi} \bigg)^{1/4} \exp \bigg( -\frac{m}{4L_1L_2L_3} \bigg\vert \int d^3 x_0 \; \chi (\vec{x}_0,y) \bigg\vert^2 \bigg) \times  \nonumber \eeq
\beq \times   \prod_{k \geq 1} \Bigg( \frac{(m^2+ \vert \vec{p}_k \vert^2)^{1/4}}{\pi^{1/2}}    \exp \bigg( -  \frac{\sqrt{m^2+ \vert \vec{p}_k \vert^2}}{L_1L_2L_3}  \bigg\vert \int d^3 x_k \; \chi (\vec{x}_k,y) e^{i \vec{p}_k \cdot \vec{x}_k} \bigg\vert ^2  \bigg) \Bigg) \Bigg] \times \nonumber  \eeq
\beq \times \bigg( \sqrt{n_0!} \sum_{C_0=0}^{\lfloor n/2 \rfloor} \frac{(-1)^{C_0} m^{\frac{n_0}{2} - C_0}}{2^{C_0} C_0! (n_0-2C_0)!} \bigg\vert \int d^3 x' \; \chi (\vec{x}',y) \bigg\vert^{n_0-2C_0} \bigg) \times \nonumber \eeq
\beq \times \prod_{k} \Bigg(   \sqrt{2^{n_k+n_{-k}} n_k! n_{-k}!}     \sum_{C_k=0}^{\min (n_k, n_{-k})} \frac{(-1)^{C_k} 2^{\frac{n_k+n_{-k}}{2}-C_k} (m^2 + \vert \vec{p}_k \vert^2)^{\frac{n_k+n_{-k}}{4} - \frac{C_k}{2}} }{2^{n_k+n_{-k}-C_k} C! (n_k-C_k)! (n_{-k} -C_k )!}  \times   \nonumber \eeq
\beq \times \bigg\vert \int d^3 x'' \; \chi (\vec{x}'',y) e^{i \vec{p}_k \cdot \vec{x}''} \bigg\vert^{n_k-2C_k} \exp \bigg(i (n_k-n_{-k}) \Im \ln \int d^3 x''' \; \chi (\vec{x}''',y) e^{i \vec{p}_k \cdot \vec{x}'''} \bigg) \bigg) \eeq
We can utilize Eq \ref{xiGeneralFinal} in order to obtain realistic interpretation of ensemble of states. In particular, the density matrix 
\beq \sum_k \bigg( C_k \vert \sharp (\vec{0}) = n_{k0}, \sharp (\vec{p}_1) = n_{k1}, \sharp (-\vec{p}_1) = n_{k,-1}, \sharp (\vec{p}_2) = n_{k2}, \sharp (- \vec{p}_2) = n_{k,-2}, \cdots \rangle \times \nonumber \eeq
 \beq \times \langle \sharp (\vec{0}) = n_{k0}, \sharp (\vec{p}_1) = n_{k1}, \sharp (-\vec{p}_1) = n_{k,-1}, \sharp (\vec{p}_2) = n_{k2}, \sharp (- \vec{p}_2) = n_{k,-2}, \cdots \vert \bigg) \eeq 
is described as 
\beq  \xi_{\sigma = \vert \cdots \rangle \langle \cdots \vert} (\vec{x},y)  \nonumber \eeq
\beq = \sum_k \bigg\{ e^{i \vec{p}_{fly} \cdot \vec{x}} \Bigg[ \bigg( \frac{m}{2 \pi} \bigg)^{1/4} \exp \bigg( -\frac{m}{4L_1L_2L_3} \bigg\vert \int d^3 x_0 \; \chi (\vec{x}_0,y) \bigg\vert^2 \bigg) \times  \nonumber \eeq
\beq \times   \prod_{k \geq 1} \Bigg( \frac{(m^2+ \vert \vec{p}_k \vert^2)^{1/4}}{\pi^{1/2}}    \exp \bigg( -  \frac{\sqrt{m^2+ \vert \vec{p}_k \vert^2}}{L_1L_2L_3}  \bigg\vert \int d^3 x_k \; \chi (\vec{x}_k,y) e^{i \vec{p}_k \cdot \vec{x}_k} \bigg\vert ^2  \bigg) \Bigg) \Bigg] \times \nonumber  \eeq
\beq \times \bigg( \sqrt{n_0!} \sum_{C_0=0}^{\lfloor n/2 \rfloor} \frac{(-1)^{C_0} m^{\frac{n_0}{2} - C_0}}{2^{C_0} C_0! (n_0-2C_0)!} \bigg\vert \int d^3 x' \; \chi (\vec{x}',y) \bigg\vert^{n_0-2C_0} \bigg) \times \nonumber \eeq
\beq \times \prod_{k} \Bigg(   \sqrt{2^{n_k+n_{-k}} n_k! n_{-k}!}     \sum_{C_k=0}^{\min (n_k, n_{-k})} \frac{(-1)^{C_k} 2^{\frac{n_k+n_{-k}}{2}-C_k} (m^2 + \vert \vec{p}_k \vert^2)^{\frac{n_k+n_{-k}}{4} - \frac{C_k}{2}} }{2^{n_k+n_{-k}-C_k} C! (n_k-C_k)! (n_{-k} -C_k )!}  \times   \nonumber \eeq
\beq \times \bigg\vert \int d^3 x'' \; \chi (\vec{x}'',y) e^{i \vec{p}_k \cdot \vec{x}''} \bigg\vert^{n_k-2C_k} \exp \bigg(i (n_k-n_{-k}) \Im \ln \int d^3 x''' \; \chi (\vec{x}''',y) e^{i \vec{p}_k \cdot \vec{x}'''} \bigg) \bigg) \bigg\} \label{DensityFinal} \eeq
We would now like to define creation and annihilation operators. However, we can no longer use the derivatives that we used in the previous section. The reason is that, as far as infinitesimal displacement is concerned, we have only one degree of freedom, namely $y$. This is not enough to define more than one partial derivative without unwanted linear dependence. The way around it is to utilize finite definition of partial derivatives as opposed to infinitesimal one; namely, for $f \colon (r_0, r_1, \theta_1, \cdots, r_n, \theta_n)\mapsto \mathbb{C}$ we define 
\beq \partial_{\theta_k}^{(\alpha)} f = \frac{\alpha^{N+ \frac{3}{2}}}{2^{1/2} \pi^{n+ \frac{1}{2}}} r_k^2 \int d^{2n+1} x' \; (\theta_k'- \theta_k) f(\vec{x}') e^{- \frac{\alpha}{2} \vert \vec{x}_n' - \vec{x}_n \vert^2} \eeq
\beq \partial_{r_k}^{(\alpha)} f= \frac{\alpha^{N+ \frac{3}{2}}}{2^{1/2} \pi^{N+ \frac{1}{2}}} \int d^{2n+1}x' \; (r_k' -r_k ) f(\vec{x}') e^{- \frac{\alpha}{2} \vert \vec{x}_n' - \vec{x}_n \vert^2} \eeq
which can be shown to approximate the corresponding derivatives in the event that $\alpha$ is so large that $f (\vec{x}')$ is approximately linear within the range where $e^{-\frac{\alpha}{2} \vert \vec{x}' - \vec{x} \vert^2}$ is far from zero.  Now we would like to replace integrals over $r_k$-s and $\theta_k$-s with the single $y$-integral where $r_k$ and $\theta_k$ are being replaced by $R^{(\chi)} (y)$ and $\Theta^{(\chi)} (y)$, respectively. First, we recall that our space is compactified, 
\beq x^1+ L_1 = x^1 \; , \; x^2 +L_2 = x^2 \; , \; x^3 +L_3 = x^3 \; , \; x^5 + L_5 = x^5 \eeq
Secondly, for any given $\phi$ we will define $\phi^{(N)}$ as a sum of its first $N$ Fourier components, 
\beq \phi^{(N)} (\vec{x}) = \frac{1}{L_1L_2L_3} \sum_{k=-N}^N \bigg[ e^{i \vec{p}_k \cdot \vec{x}} \bigg( \int d^3 x' \; \phi (\vec{x}') e^{-i \vec{p}_k \cdot \vec{x}'} \bigg) \bigg] \label{PhiN} \eeq
and, finally, we will assume that $\chi$ behaves in such a way that $\{\chi_y^{(N)} \vert 0 \leq y < L_5\}$ is distributed in $\mathbb{R}^{2N+1}$ with probability density $\rho$. In this case, the integral over $\phi^{(N)}$ will be replaced with an integral over $y$ via the following scheme: 
\beq \int d^{2N+1} \phi^{(N)} \; f(\phi^{(N)}) \longrightarrow \frac{1}{L_5} \int dy' \; \frac{f(\chi^{(N)} (y'))}{\rho (\chi^{(N)} (y'))} \label{IntegralConversion} \eeq
from which we read off the following definitions of partial derivatives: 
\beq ({\cal D}^{(N, \chi, \rho (\phi))}_{\Theta_k} \xi) (\vec{x},y) =  \frac{\alpha^{N+ \frac{3}{2}}}{2^{1/2} \pi^{N+ \frac{1}{2}}L_5} R_k^2 (\chi_y) \times \nonumber \eeq
\beq \times \int d y' \; \frac{( \Theta_k (\chi^{(N)}_{y'}) - \Theta_k (\chi^{(N)}_{y})) \xi(\vec{x},y') e^{- \frac{\alpha}{2} \vert \chi_{y'} - \chi_y \vert^2}}{\rho (\chi^{(N)} (y'))} \label{DThetaShort} \eeq
\beq  ({\cal D}^{(N, \chi, \rho (\phi))}_{R_k} \xi) (\vec{x},y) = \frac{\alpha^{N+ \frac{3}{2}}}{2^{1/2} \pi^{N+ \frac{1}{2}} L_5} \int dy' \; \frac{(R_k (\chi^{(N)}_{y'}) -R_k (\chi_y^{(N)}) ) \xi(\vec{x},y') e^{- \frac{\alpha}{2} \vert \chi_{y'} - \chi_y \vert^2}}{\rho (\chi^{(N)} (y'))} \label{dRShort} \eeq
\beq  ({\cal D}^{(N, \chi, \rho (\phi))}_{R_0} \xi) (\vec{x},y) = \frac{\alpha^{N+ \frac{3}{2}}}{2^{1/2} \pi^{N+ \frac{1}{2}} L_5} \int dy' \; \frac{(R_0 (\chi^{(N)}_{y'}) -R_0 (\chi_y^{(N)}) ) \xi(\vec{x},y') e^{- \frac{\alpha}{2} \vert \chi_{y'} - \chi_y \vert^2}}{\rho (\chi^{(N)} (y'))} \label{dR0Short} \eeq
Let us now substitute explicit expressions for $R$ and $\Theta$ in order to come up with an expression that only involves $\chi$ and $\xi$, however complicated that might be. First of all, one can easily show that 
\beq 0 \leq k \leq N \Longrightarrow R_k (\phi) = R_k (\phi^{(N)}) \eeq
\beq 1 \leq k \leq N \Longrightarrow \Theta_k (\phi) = \Theta_k (\phi^{(N)}) \eeq
and, therefore, 
\beq R_0 (\phi^{(N)}) = R_{0} (\phi) = \frac{1}{\sqrt{L_1L_2L_3}} \bigg\vert \int d^3 x \; \phi (\vec{x})   \bigg\vert \eeq
\beq  1 \leq k \leq N \Longrightarrow R_k (\phi^{(N)}) = R_k (\phi)  = \sqrt{\frac{2}{L_1L_2L_3}} \bigg\vert \int d^3 x \; \phi (\vec{x}) e^{i \vec{p}_k \cdot \vec{x}} \bigg\vert  \eeq
\beq 1 \leq k \leq N \Longrightarrow \Theta_k (\phi^{(N)}) = \Theta_k (\phi) =  \Im \ln \int \phi (\vec{x}) e^{i \vec{p}_k \cdot \vec{x}} d^3 x \eeq
We will then convert $R$ and $\Theta$ into functions of $y$ as follows: 
\beq R_0^{(\chi)} (y) = R_0 (\chi_y) = \frac{1}{\sqrt{L_1L_2L_3}} \bigg\vert \int d^3 x \; \chi_y (\vec{x}) \bigg\vert = \frac{1}{\sqrt{L_1L_2L_3}} \bigg\vert \int d^3 x \; \chi_y (\vec{x},y) \bigg\vert \label{R0(y)} \eeq
\beq k \neq 0 \Longrightarrow R^{(\chi)}_k (y) = R_k (\chi_y) = \sqrt{\frac{2}{L_1L_2L_3}} \bigg\vert \int d^3 x \; \chi_y (\vec{x}) e^{i \vec{p}_k \cdot \vec{x}} \bigg\vert = \nonumber \eeq
\beq = \sqrt{\frac{2}{L_1L_2L_3}} \bigg\vert \int d^3 x \; \chi (\vec{x},y) e^{i \vec{p}_k \cdot \vec{x}} \bigg\vert \label{Rk(y)} \eeq
\beq \Theta_k^{(\chi)} (y) = \Theta_k (\chi_y) = \Im \ln \int d^3 x \; \chi_y (\vec{x}) e^{i \vec{p}_k \cdot \vec{x}} = \Im \ln \int d^3 x \; \chi (\vec{x},y) e^{i \vec{p}_k \cdot \vec{x}} \label{Thetak(y)} \eeq
Furthermore, we will assume that the probability distribution $\rho$ is Gaussian, 
\beq \rho^{(\beta,N)} (\phi) =  \bigg(\frac{\beta}{2 \pi} \bigg)^{N+ \frac{1}{2}}  \exp \bigg(- \frac{\beta}{2L_1L_2L_3} \sum_{k=-N}^N \bigg\vert \int d^3 x' \; \phi (\vec{x}') e^{-i \vec{p}_k \cdot \vec{x}'} \bigg\vert^2 \bigg) \eeq
From this we define
\beq \rho^{(\beta, N, \chi)} (y) = \rho^{(\beta,N)} (\chi_y) = \bigg(\frac{\beta}{2 \pi} \bigg)^{N+ \frac{1}{2}}  \exp \bigg(- \frac{\beta}{2L_1L_2L_3} \sum_{k=-N}^N \bigg\vert \int d^3 x' \; \chi_y (\vec{x}') e^{-i \vec{p}_k \cdot \vec{x}'} \bigg\vert^2 \bigg) = \nonumber \eeq
\beq =  \bigg(\frac{\beta}{2 \pi} \bigg)^{N+ \frac{1}{2}}  \exp \bigg(- \frac{\beta}{2L_1L_2L_3} \sum_{k=-N}^N \bigg\vert \int d^3 x' \; \chi (\vec{x}',y) e^{-i \vec{p}_k \cdot \vec{x}'} \bigg\vert^2 \bigg)  \label{rho} \eeq
Now, by plugging in Eq \ref{Rk(y)}, \ref{Thetak(y)} and \ref{rho} into Eq \ref{DThetaShort}, we obtain
\beq ({\cal D}^{(N, \chi, \alpha, \beta)}_{\Theta_k} \xi) (\vec{x},y) =  \frac{2^{N+1}  \alpha^{N+ \frac{3}{2}}}{  \beta^{N+\frac{1}{2}} L_5 L_1L_2L_3}\bigg\vert \int d^3 x' \; \chi (\vec{x}',y) e^{i \vec{p}_k \cdot \vec{x}'} \bigg\vert^2  \times \nonumber \eeq
\beq \times \int d y' \; \bigg[ \bigg( \Im \ln \int \chi (\vec{x}'',y') e^{i \vec{p}_k \cdot \vec{x}''} d^3 x''  -  \Im \ln \int d^3 x''' \; \chi (\vec{x}''',y) e^{i \vec{p}_k \cdot \vec{x}'''}  \bigg) \times \nonumber \eeq
\beq \times \xi(\vec{x},y')  \exp \bigg(- \frac{\alpha}{2L_1L_2L_3} \sum_{k=-N}^N \bigg\vert \int d^3 x'''' \; \chi (\vec{x}'''',y') e^{-i \vec{p}_k \cdot \vec{x}''''} \bigg\vert^2 \bigg) \nonumber \eeq
\beq \times \exp \bigg(\frac{\beta}{2L_1L_2L_3} \sum_{k=-N}^N \bigg\vert \int d^3 x''''' \; \chi (\vec{x}''''',y') e^{-i \vec{p}_k \cdot \vec{x}'''''} \bigg\vert^2 \bigg) \bigg]  \label{dThetakLong} \eeq
On the other hand, if we plug in  Eq \ref{Rk(y)}, \ref{Thetak(y)} and \ref{rho} into Eq \ref{dRShort}, we obtain
\beq k \neq 0 \Longrightarrow ({\cal D}^{(N, \chi, \alpha, \beta)}_{R_k} \xi) (\vec{x},y) =  \frac{2^{N+\frac{1}{2}}  \alpha^{N+ \frac{3}{2}}}{  \beta^{N+\frac{1}{2}} L_5 \sqrt{L_1L_2L_3}}  \times \nonumber \eeq
\beq \times \int d y' \; \bigg[ \bigg(\bigg\vert \int d^3 x' \; \chi (\vec{x}',y) e^{i \vec{p}_k \cdot \vec{x}'}\bigg\vert  -  \bigg\vert \int d^3 x'' \; \chi (\vec{x}'',y) e^{i \vec{p}_k \cdot \vec{x}''} \bigg\vert \bigg) \times \nonumber \eeq
\beq \times \xi(\vec{x},y')  \exp \bigg(- \frac{\alpha}{2L_1L_2L_3} \sum_{k=-N}^N \bigg\vert \int d^3 x''' \; \chi (\vec{x}''',y') e^{-i \vec{p}_k \cdot \vec{x}'''} \bigg\vert^2 \bigg) \nonumber \eeq
\beq \times \exp \bigg(\frac{\beta}{2L_1L_2L_3} \sum_{k=-N}^N \bigg\vert \int d^3 x'''' \; \chi (\vec{x}'''',y') e^{-i \vec{p}_k \cdot \vec{x}''''} \bigg\vert^2 \bigg) \bigg]  \label{dRkLong} \eeq
Finally, if we plug in Eq \ref{R0(y)} and \ref{rho} into Eq \ref{dR0Short}, we obtain
\beq ({\cal D}^{(N, \chi, \alpha, \beta)}_{R_0} \xi) (\vec{x},y) =  \frac{2^N  \alpha^{N+ \frac{3}{2}}}{  \beta^{N+\frac{1}{2}} L_5 \sqrt{L_1L_2L_3}}  \times \nonumber \eeq
\beq \times \int d y' \; \bigg[ \bigg(\bigg\vert \int d^3 x' \; \chi (\vec{x}',y) e^{i \vec{p}_0 \cdot \vec{x}'} \bigg\vert  -  \bigg\vert \int d^3 x'' \; \chi (\vec{x}'',y) e^{i \vec{p}_0 \cdot \vec{x}''} \bigg\vert \bigg) \times \nonumber \eeq
\beq \times \xi(\vec{x},y')  \exp \bigg(- \frac{\alpha}{2L_1L_2L_3} \sum_{k=-N}^N \bigg\vert \int d^3 x''' \; \chi (\vec{x}''',y') e^{-i \vec{p}_k \cdot \vec{x}'''} \bigg\vert^2 \bigg) \nonumber \eeq
\beq \times \exp \bigg(\frac{\beta}{2L_1L_2L_3} \sum_{k=-N}^N \bigg\vert \int d^3 x'''' \; \chi (\vec{x}'''',y') e^{-i \vec{p}_k \cdot \vec{x}''''} \bigg\vert^2 \bigg) \bigg] \label{dR0Long}  \eeq
Now that we have defined the derivatives, we are going to use them to define creation and annihilation operators. By looking at Eq \ref{AnnihilationInfiniteOriginal} and \ref{CreationInfiniteOriginal}, and making appropriate substitutions, we obtain
\beq [a_{p_k}^{(N, \chi, \alpha, \beta)} (\xi)] (\vec{x},y) = \frac{e^{-i \Theta_k (\chi_y)}}{2} \bigg( R_k (\chi_y) \xi (\vec{x},y) (m^2 + \vert \vec{p}_k \vert^2)^{1/4} + \nonumber \eeq
\beq + \frac{1}{ (m^2 + \vert \vec{p}_k \vert^2)^{1/4}} ({\cal D}^{(N, \chi, \alpha, \beta)}_{R_k} \xi)(\vec{x},y) - \frac{i}{R_k (\phi)  (m^2 + \vert \vec{p}_k \vert^2)^{1/4}} ({\cal D}^{(N, \chi, \alpha, \beta)}_{\Theta_k} \xi) (\vec{x},y)\bigg)  \label{Annihilation5Short} \eeq
\beq [a_{p_k}^{\dagger (N, \chi, \alpha, \beta)} (\xi)] (\vec{x},y) = \frac{e^{i \Theta_k (\chi_y)}}{2} \bigg( R_k (\chi_y) \xi (\vec{x},y) (m^2 + \vert \vec{p}_k \vert^2)^{1/4} - \nonumber \eeq
\beq - \frac{1}{ (m^2 + \vert \vec{p}_k \vert^2)^{1/4}} ({\cal D}^{(N, \chi, \alpha, \beta)}_{R_k} \xi)(\vec{x},y) - \frac{i}{R_k (\phi)  (m^2 + \vert \vec{p}_k \vert^2)^{1/4}} ({\cal D}^{(N, \chi, \alpha, \beta)}_{\Theta_k} \xi) (\vec{x},y)\bigg)  \label{Creation5Short} \eeq
If we plug in Eq \ref{Rk(y)}, \ref{dRkLong} and \ref{dThetakLong} into Eq \ref{Annihilation5Short}, we obtain
\beq [a_{p_k}^{(N, \alpha, \beta, \chi)} (\xi)] (\vec{x},y) = \frac{1}{2} \bigg[ \exp \bigg( - \ii \; \Im \ln \int d^3 x' \; \chi (\vec{x}',y) e^{i \vec{p}_k \cdot \vec{x}'} \bigg) \bigg] \times \nonumber \eeq 
\beq \times \Bigg\{ \xi (\vec{x},y) (m^2 + \vert \vec{p}_k \vert^2)^{1/4} \bigg( \sqrt{\frac{2}{L_1L_2L_3}} \bigg\vert \int d^3 x'' \; \chi (\vec{x}'',y) e^{i \vec{p}_k \cdot \vec{x}''} \bigg\vert \bigg) + \nonumber \eeq
\beq + \frac{1}{ (m^2 + \vert \vec{p}_k \vert^2)^{1/4}} \bigg\{ \frac{2^{N+\frac{1}{2}}  \alpha^{N+ \frac{3}{2}}}{  \beta^{N+\frac{1}{2}} L_5 \sqrt{L_1L_2L_3}}  \times \nonumber \eeq
\beq \times \int d y' \; \bigg[ \bigg(\bigg\vert \int d^3 x''' \; \chi (\vec{x}''',y) e^{i \vec{p}_k \cdot \vec{x}'''} \bigg\vert  - \bigg\vert \int d^3 x'''' \; \chi (\vec{x}'''',y) e^{i \vec{p}_k \cdot \vec{x}''''} \bigg\vert \bigg) \times \nonumber \eeq
\beq \times \xi(\vec{x},y')  \exp \bigg(- \frac{\alpha}{2L_1L_2L_3} \sum_{k=-N}^N \bigg\vert \int d^3 x''''' \; \chi (\vec{x}''''',y') e^{-i \vec{p}_k \cdot \vec{x}'''''} \bigg\vert^2 \bigg) \nonumber \eeq
\beq \times \exp \bigg(\frac{\beta}{2L_1L_2L_3} \sum_{k=-N}^N \bigg\vert \int d^3 x'''''' \; \chi (\vec{x}'''''',y') e^{-i \vec{p}_k \cdot \vec{x}''''''} \bigg\vert^2 \bigg) \bigg] \bigg\} - \nonumber \eeq
\beq - \frac{i}{ (m^2 + \vert \vec{p}_k \vert^2)^{1/4} }  \frac{2^{N+\frac{1}{2}}  \alpha^{N+ \frac{3}{2}}}{  \beta^{N+\frac{1}{2}} L_5 \sqrt{L_1L_2L_3}}\bigg\vert \int d^3 x''''''' \; \chi (\vec{x}''''''',y) e^{i \vec{p}_k \cdot \vec{x}'''''''} \bigg\vert \times \nonumber \eeq
\beq \times \int d y' \; \bigg[ \bigg( \Im \ln \int \chi (\vec{x}'''''''',y') e^{i \vec{p}_k \cdot \vec{x}''''''''} d^3 x''''''''  -  \Im \ln \int d^3 x''''''''' \; \chi (\vec{x}''''''''',y) e^{i \vec{p}_k \cdot \vec{x}'''''''''}  \bigg) \times \nonumber \eeq
\beq \times \xi(\vec{x},y')  \exp \bigg(- \frac{\alpha}{2L_1L_2L_3} \sum_{k=-N}^N \bigg\vert \int d^3 x'''''''''' \; \chi (\vec{x}'''''''''',y') e^{-i \vec{p}_k \cdot \vec{x}''''''''''} \bigg\vert^2 \bigg) \nonumber \eeq
\beq \times \exp \bigg(\frac{\beta}{2L_1L_2L_3} \sum_{k=-N}^N \bigg\vert \int d^3 x''''''''''' \; \chi (\vec{x}''''''''''',y') e^{-i \vec{p}_k \cdot \vec{x}'''''''''''} \bigg\vert^2 \bigg) \bigg]  \Bigg\}  \label{AnnihilateFinal} \eeq
On the other hand, if we plug in Eq \ref{Rk(y)}, \ref{dRkLong} and \ref{dThetakLong} into Eq \ref{Creation5Short}, we obtain
\beq [a_{p_k}^{\dagger (N, \alpha, \beta, \chi)} (\xi)] (\vec{x},y) = \frac{1}{2} \bigg[ \exp \bigg(  \ii \; \Im \ln \int d^3 x' \; \chi (\vec{x}',y) e^{i \vec{p}_k \cdot \vec{x}'} \bigg) \bigg] \times \nonumber \eeq 
\beq \times \Bigg\{ \xi (\vec{x},y) (m^2 + \vert \vec{p}_k \vert^2)^{1/4} \bigg( \sqrt{\frac{2}{L_1L_2L_3}} \bigg\vert \int d^3 x'' \; \chi (\vec{x}'',y) e^{i \vec{p}_k \cdot \vec{x}''} \bigg\vert \bigg) - \nonumber \eeq
\beq - \frac{1}{ (m^2 + \vert \vec{p}_k \vert^2)^{1/4}} \bigg\{ \frac{2^{N+\frac{1}{2}}  \alpha^{N+ \frac{3}{2}}}{  \beta^{N+\frac{1}{2}} L_5 \sqrt{L_1L_2L_3}}  \times \nonumber \eeq
\beq \times \int d y' \; \bigg[ \bigg(\bigg\vert \int d^3 x''' \; \chi (\vec{x}''',y) e^{i \vec{p}_k \cdot \vec{x}'''}  \bigg\vert -  \bigg\vert \int d^3 x'''' \; \chi (\vec{x}'''',y) e^{i \vec{p}_k \cdot \vec{x}''''} \bigg\vert \bigg) \times \nonumber \eeq
\beq \times \xi(\vec{x},y')  \exp \bigg(- \frac{\alpha}{2L_1L_2L_3} \sum_{k=-N}^N \bigg\vert \int d^3 x''''' \; \chi (\vec{x}''''',y') e^{-i \vec{p}_k \cdot \vec{x}'''''} \bigg\vert^2 \bigg) \nonumber \eeq
\beq \times \exp \bigg(\frac{\beta}{2L_1L_2L_3} \sum_{k=-N}^N \bigg\vert \int d^3 x'''''' \; \chi (\vec{x}'''''',y') e^{-i \vec{p}_k \cdot \vec{x}''''''} \bigg\vert^2 \bigg) \bigg] \bigg\} - \nonumber \eeq
\beq - \frac{i}{ (m^2 + \vert \vec{p}_k \vert^2)^{1/4} }  \frac{2^{N+\frac{1}{2}}  \alpha^{N+ \frac{3}{2}}}{  \beta^{N+\frac{1}{2}} L_5 \sqrt{L_1L_2L_3}}\bigg\vert \int d^3 x''''''' \; \chi (\vec{x}''''''',y) e^{i \vec{p}_k \cdot \vec{x}'''''''} \bigg\vert  \times \nonumber \eeq
\beq \times \int d y' \; \bigg[ \bigg( \Im \ln \int \chi (\vec{x}'''''''',y') e^{i \vec{p}_k \cdot \vec{x}''''''''} d^3 x''''''''  -  \Im \ln \int d^3 x''''''''' \; \chi (\vec{x}''''''''',y) e^{i \vec{p}_k \cdot \vec{x}'''''''''}  \bigg) \times \nonumber \eeq
\beq \times \xi(\vec{x},y')  \exp \bigg(- \frac{\alpha}{2L_1L_2L_3} \sum_{k=-N}^N \bigg\vert \int d^3 x'''''''''' \; \chi (\vec{x}'''''''''',y') e^{-i \vec{p}_k \cdot \vec{x}''''''''''} \bigg\vert^2 \bigg) \nonumber \eeq
\beq \times \exp \bigg(\frac{\beta}{2L_1L_2L_3} \sum_{k=-N}^N \bigg\vert \int d^3 x''''''''''' \; \chi (\vec{x}''''''''''',y') e^{-i \vec{p}_k \cdot \vec{x}'''''''''''} \bigg\vert^2 \bigg) \bigg]  \Bigg\}  \label{CreateFinal} \eeq
The 2D oscillator that we were using covers non-zero momentum. On the other hand, the zero momentum needs to be done separately by using 1D oscillator. By looking at Eq \ref{CreatAnnihilZeroCont} and making appropriate substitutions, we obtain
\beq (a^{(N, \chi, \alpha, \beta)}_0 \xi) (\vec{x},y)= \frac{\sqrt{m}}{2} R_0 (\chi_y) \xi (\vec{x},y)  + \frac{1}{\sqrt{m}} ({\cal D}^{(N, \chi, \alpha, \beta)}_{R_0} \xi) (\vec{x},y) \eeq
\beq (a_0^{\dagger(N, \chi, \alpha, \beta)} \xi) (\vec{x},y)= \frac{\sqrt{m}}{2} R_0 (\chi_y) \xi (\vec{x},y)- \frac{1}{\sqrt{m}} ({\cal D}^{(N, \chi, \alpha, \beta)}_{R_0} \xi) (\vec{x},y) \eeq
By substituting Eq \ref{R0(y)} and \ref{dR0Long} this becomes 
\beq (a^{(N, \chi, \alpha, \beta)}_0 \xi) (\vec{x},y)= \frac{1}{2} \sqrt{\frac{m}{L_1L_2L_3}} \xi (\vec{x},y)   \bigg\vert \int d^3 x \; \chi (\vec{x},y)   \bigg\vert  + \nonumber \eeq
\beq +    \frac{2^N  \alpha^{N+ \frac{3}{2}}}{  \beta^{N+\frac{1}{2}} L_5 \sqrt{m L_1L_2L_3}}  \int d y' \; \bigg[ \bigg(\bigg\vert \int d^3 x' \; \chi (\vec{x}',y) e^{i \vec{p}_0 \cdot \vec{x}'} \bigg\vert  -  \bigg\vert \int d^3 x'' \; \chi (\vec{x}'',y) e^{i \vec{p}_0 \cdot \vec{x}''} \bigg\vert \bigg) \times \nonumber \eeq
\beq \times \xi(\vec{x},y')  \exp \bigg(- \frac{\alpha}{2L_1L_2L_3} \sum_{k=-N}^N \bigg\vert \int d^3 x''' \; \chi (\vec{x}''',y') e^{-i \vec{p}_k \cdot \vec{x}'''} \bigg\vert^2 \bigg) \nonumber \eeq
\beq \times \exp \bigg(\frac{\beta}{2L_1L_2L_3} \sum_{k=-N}^N \bigg\vert \int d^3 x'''' \; \chi (\vec{x}'''',y') e^{-i \vec{p}_k \cdot \vec{x}''''} \bigg\vert^2 \bigg) \bigg]  \label{Annihilate0Final} \eeq

\beq (a_0^{\dagger (N, \chi, \alpha, \beta)} \xi) (\vec{x},y)= \frac{1}{2} \sqrt{\frac{m}{L_1L_2L_3}} \xi (\vec{x},y)   \bigg\vert \int d^3 x \; \chi (\vec{x},y)   \bigg\vert  - \nonumber \eeq
\beq -    \frac{2^N  \alpha^{N+ \frac{3}{2}}}{  \beta^{N+\frac{1}{2}} L_5 \sqrt{m L_1L_2L_3}}  \int d y' \; \bigg[ \bigg(\bigg\vert \int d^3 x' \; \chi (\vec{x}',y) e^{i \vec{p}_0 \cdot \vec{x}'} \bigg\vert  -  \bigg\vert \int d^3 x'' \; \chi (\vec{x}'',y) e^{i \vec{p}_0 \cdot \vec{x}''} \bigg\vert \bigg) \times \nonumber \eeq
\beq \times \xi(\vec{x},y')  \exp \bigg(- \frac{\alpha}{2L_1L_2L_3} \sum_{k=-N}^N \bigg\vert \int d^3 x''' \; \chi (\vec{x}''',y') e^{-i \vec{p}_k \cdot \vec{x}'''} \bigg\vert^2 \bigg) \nonumber \eeq
\beq \times \exp \bigg(\frac{\beta}{2L_1L_2L_3} \sum_{k=-N}^N \bigg\vert \int d^3 x'''' \; \chi (\vec{x}'''',y') e^{-i \vec{p}_k \cdot \vec{x}''''} \bigg\vert^2 \bigg) \bigg] \label{Create0Final} \eeq

\subsection*{7. Dynamics of $\xi (\vec{x},y,t)$} 

So far we have just given the kinematical definitions of quantum states. Let us now describe the dynamics. The immediate question the reader will have is that we already did the dynamics, in Chapter 2. The reason we have to do it again is that the equivalence between Feynmann path integral formulation and Fock space formulation holds only in case of \emph{true} functionals as opposed to coarse-grained ones discussed in this paper. Since we are hypothesizing that the nature operates on coarse-grained functionals, the Feynmann path integral formalism and Fock space formalism are the same only up to certain approximation. Therefore, if we are interested in exact theory it would have to be \emph{either} based on Feynmann path integral \emph{or} based on Fock space, not both. I, personally, favor the Feynmann path integral version of the theory simply because -- as a reader will find -- it is a lot simpler. However, the point of this paper is to show how \emph{both} formalisms can be translated into our framework. Therefore, in order to make this paper complete, we have to translate both formalisms into two separate theories, at most one of which can be true -- and leave it up to the reader to decide. So let us go ahead and translate Fock-space-based theory into our framework. 

Even though we are doing Fock space, we will still take some insights from Feynmann path integral, and simply use them in a different context. We recall that, in quantum mechanics case, path integral can be produced from 
\beq \psi (\vec{x},t) = \int d^3x' \; \psi (\vec{x}', t- \delta t) \exp \bigg( i \bigg\vert \frac{\vec{x}-\vec{x}'}{\delta t} \bigg\vert^2 - i V (x) \bigg) \label{QuantumMechanicsPathIntegral} \eeq 
We will now assume the preferred time and, therefore, the Lagrangian above is analogous to the integral of $\cal L$ over spacelike hypersurface,
\beq S (\phi; t) = \int d^3 x \; {\cal L} (\phi; \vec{x}, t) \eeq
From this, we read off the QFT version of Eq \ref{QuantumMechanicsPathIntegral} as 
\beq \psi (\phi^{(N)},t) = \int {\cal D} \phi'^{(N)} \; \bigg\{ \psi (\phi'^{(N)}, t- \delta t)  \times \nonumber \eeq
\beq \times \exp \bigg[ i \int d^3 x \bigg( \frac{1}{2} \bigg(\frac{\phi^{(N)} (\vec{x}) - \phi'^{(N)}) (\vec{x})}{\delta t} \bigg)^2  - \frac{m^2}{2} (\phi^{(N)} (\vec{x}))^2 - \frac{\lambda}{4} (\phi^{(N)} (\vec{x}))^4 \bigg) \bigg] \bigg\} \label{Hypersurfaces1} \eeq 
One should note that we used $\phi^{(N)}$ instead of $\phi$. The reason for this is that, if we were to use $\phi$ we would get infinitely many contributions from arbitrarily high momenta, leading to intractable results. The purpose of $N$ is the same as the purpose of ultraviolet cutoff $\Lambda$ in QFT calculations. On the first glance, one might think that since we plan to substitute integration over $\phi$ with integration over $y$ per Eq \ref{IntegralConversion} the theory would be well defined even with $\phi$ being used instead of $\phi^{(N)}$. However,  Eq \ref{IntegralConversion} includes $\rho^{(N,\chi, \beta)}$ and, as notation implies, we still need to know $N$ in order to know $\rho$. If $N$ were infinite then $\rho$ would have been infinitesimal, leading to mathematical ambiguities. In any case, Eq \ref{PhiN} tells us 
\beq \phi^{(N)} (\vec{x}) = \frac{1}{L_1L_2L_3} \sum_{k=-N}^N \bigg( e^{i \vec{p}_k \cdot \vec{x}} \bigg( \int d^3 x' \; \phi (\vec{x}') e^{-i \vec{p}_k \cdot \vec{x}'} \bigg) \bigg) \label{(N)1} \eeq
and, therefore 
\beq \frac{\phi^{(N)} (\vec{x}) - \phi'^{(N)} (\vec{x})}{\delta t} = \frac{1}{L_1L_2L_3} \sum_{k=-N}^N \bigg( e^{i \vec{p}_k \cdot \vec{x}} \bigg( \int d^3 x' \; \frac{\phi (\vec{x}') - \phi' (\vec{x}')}{\delta t} e^{-i \vec{p}_k \cdot \vec{x}'} \bigg) \bigg) \label{(N)2} \eeq
by substituting Eq \ref{(N)1} and \ref{(N)2} into Eq \ref{Hypersurfaces1} we obtain
\beq \psi (\phi,t) = \int {\cal D} \phi' \; \bigg\{ \psi (\phi', t- \delta t)  \times \nonumber \eeq
\beq \times \exp \bigg[ i \int d^3 x \bigg( \frac{1}{2L_1^2L_2^2L_3^2} \bigg(\sum_{k=-N}^N \bigg( e^{i \vec{p}_k \cdot \vec{x}}  \int d^3 x' \; \frac{\phi (\vec{x}') - \phi' (\vec{x}')}{\delta t} e^{-i \vec{p}_k \cdot \vec{x}'}  \bigg)\bigg)^2  - \nonumber \eeq
\beq -  \frac{m^2}{2L_1^2L_2^2L_3^2} \bigg(\sum_{k=-N}^N \bigg( e^{i \vec{p}_k \cdot \vec{x}} \bigg( \int d^3 x' \; \phi (\vec{x}') e^{-i \vec{p}_k \cdot \vec{x}'} \bigg) \bigg) \bigg)^2 - \nonumber \eeq
\beq -  \frac{\lambda}{4L_1^4L_2^4L_3^4} \bigg(\sum_{k=-N}^N \bigg( e^{i \vec{p}_k \cdot \vec{x}} \bigg( \int d^3 x' \; \phi (\vec{x}') e^{-i \vec{p}_k \cdot \vec{x}'} \bigg) \bigg) \bigg)^4  \bigg) \bigg] \bigg\} \eeq 
We define operation "truth value", denoted by $T$, as 
\beq T (True) = 1 \; , \; T (False) = 0 \eeq
with this notation, after the evaluating outside integral, we obtain 
\beq \psi (\phi,t) = \int {\cal D} \phi' \; \bigg\{ \psi (\phi', t- \delta t)  \times \nonumber \eeq
\beq \times \exp \bigg[ i \bigg( \frac{1}{2} \sum_{k_1=-N}^N \sum_{k_2=-N}^N \bigg( T(\vec{p}_{k_1} + \vec{p}_{k_2} = \vec{0})  \times \nonumber \eeq
\beq \times  \bigg( \int d^3 x' \; \frac{\phi (\vec{x}') - \phi' (\vec{x}')}{\delta t} e^{-i \vec{p}_1 \cdot \vec{x}'}  \bigg) \bigg( \int d^3 x' \; \frac{\phi (\vec{x}') - \phi' (\vec{x}')}{\delta t} e^{-i \vec{p}_2 \cdot \vec{x}'}  \bigg) \bigg) - \nonumber \eeq
\beq - \frac{m^2}{2} \sum_{k_1=-N}^N \sum_{k_2=-N}^N \bigg( T(\vec{p}_{k_1} + \vec{p}_{k_2} = \vec{0})  \bigg( \int d^3 x \;  \phi (\vec{x}')  e^{-i \vec{p}_{k_1} \cdot \vec{x}'}  \bigg) \bigg( \int d^3 x'' \; \phi (\vec{x}'') e^{-i \vec{p}_{k_2} \cdot \vec{x}''}  \bigg) \bigg) - \nonumber \eeq
\beq  - \frac{\lambda}{4} \bigg(\sum_{k_1=-N}^N \sum_{k_2=-N}^N \sum_{k_3=-N}^N \sum_{k_4=-N}^N \bigg( T (\vec{k_1} + \vec{k_2} + \vec{k_3} + \vec{k_4} = \vec{0})  \times \nonumber \eeq
\beq \times \bigg( \int d^3 x' \; \phi (\vec{x}') e^{-i \vec{p}_{k_1} \cdot \vec{x}'} \bigg) \bigg( \int d^3 x'' \; \phi (\vec{x}'') e^{-i \vec{p}_{k_2} \cdot \vec{x}''} \bigg) \times \nonumber \eeq 
\beq \times \bigg( \int d^3 x' \; \phi (\vec{x}') e^{-i \vec{p}_{k_3} \cdot \vec{x}'} \bigg) \times \bigg( \int d^3 x'' \; \phi (\vec{x}'') e^{-i \vec{p}_{k_4} \cdot \vec{x}''} \bigg) \bigg) \bigg)^4  \bigg) \bigg] \bigg\} \eeq 
The use of $T(\vec{p}_i + \vec{p}_j = \vec{0})$ instead of $T(\vec{p}_i = \vec{p}_j)$ might seem perplexing on the physical grounds, since it seems to imply that the momentum of a particle would change its sign without any outside interactions. However, if we recall that we are dealing with \emph{real} scalar field as opposed to complex one, it is symmetric when it comes to $\vec{p} \leftrightarrow - \vec{p}$ (which, again, underscores the fact that our geometry is Aristotelian rather than Lorentzian) and, therefore, we can use $T(\vec{p}_i + \vec{p}_j = \vec{0})$ and $T(\vec{p}_i = \vec{p}_j)$ interchangeably. 

We are now ready to convert the integral over $\phi$ into the integral over $y$. Eq \ref{IntegralConversion} tells us that the prescription of such conversion is
\beq \int d^{2N+1} \phi^{(N)} \; f(\phi^{(N)}) \longrightarrow \frac{1}{L_5} \int dy' \; \frac{f(\chi^{(N)} (y'))}{\rho (\chi^{(N)} (y'))}  \eeq
Therefore, we read off 
\beq \xi (\vec{x},y,t) = \frac{1}{L_5} \int \frac{dy'}{\rho^{(N, \chi, \beta)} (\chi^{(N)} (y'))} \; \bigg\{ \xi (\vec{x},y', t- \delta t)  \times \nonumber \eeq
\beq \times \exp \bigg[ i \bigg( \frac{1}{2} \sum_{k_1=-N}^N \sum_{k_2=-N}^N \bigg( T(\vec{p}_{k_1} + \vec{p}_{k_2} = \vec{0})  \times \nonumber \eeq
\beq \times  \bigg( \int d^3 x' \; \frac{\chi (\vec{x}',y) - \chi (\vec{x}',y')}{\delta t} e^{-i \vec{p}_1 \cdot \vec{x}'}  \bigg) \bigg( \int d^3 x'' \; \frac{\chi (\vec{x}'',y) - \chi (\vec{x}'',y')}{\delta t} e^{-i \vec{p}_2 \cdot \vec{x}''}  \bigg) \bigg) - \nonumber \eeq
\beq - \frac{m^2}{2} \sum_{k_1=-N}^N \sum_{k_2=-N}^N \bigg( T(\vec{p}_{k_1} + \vec{p}_{k_2} = \vec{0})  \bigg( \int d^3 x' \;  \chi (\vec{x}',y)  e^{-i \vec{p}_{k_1} \cdot \vec{x}'}  \bigg) \bigg( \int d^3 x''\; \chi (\vec{x}'',y) e^{-i \vec{p}_{k_2} \cdot \vec{x}''}  \bigg) \bigg) - \nonumber \eeq
\beq  - \frac{\lambda}{4} \bigg(\sum_{k_1=-N}^N \sum_{k_2=-N}^N \sum_{k_3=-N}^N \sum_{k_4=-N}^N \bigg( T (\vec{k_1} + \vec{k_2} + \vec{k_3} + \vec{k_4} = \vec{0})  \times \nonumber \eeq
\beq \times \bigg( \int d^3 x' \; \chi (\vec{x}',y) e^{-i \vec{p}_{k_1} \cdot \vec{x}'} \bigg) \bigg( \int d^3 x'' \; \chi (\vec{x}'',y) e^{-i \vec{p}_{k_2} \cdot \vec{x}''} \bigg) \times \nonumber \eeq 
\beq \times \bigg( \int d^3 x''' \; \chi (\vec{x}''',y) e^{-i \vec{p}_{k_3} \cdot \vec{x}'''} \bigg) \times \bigg( \int d^3 x'''' \; \chi (\vec{x}'''',y) e^{-i \vec{p}_{k_4} \cdot \vec{x}''''} \bigg) \bigg) \bigg)^4  \bigg) \bigg] \bigg\} \label{StepPreFinal}  \eeq 
Now Eq \ref{rho} tells us that 
\beq \rho^{(N, \chi, \beta)} =  \bigg(\frac{\beta}{2 \pi} \bigg)^{N+ \frac{1}{2}}  \exp \bigg(- \frac{\beta}{2L_1L_2L_3} \sum_{k=-N}^N \bigg\vert \int d^3 x' \; \chi (\vec{x}',y) e^{-i \vec{p}_k \cdot \vec{x}'} \bigg\vert^2 \bigg) \eeq
and, therefore Eq \ref{StepPreFinal} becomes
\beq \xi (\vec{x},y,t) = \frac{1}{L_5} \int dy' \;  \bigg\{ \bigg[  \bigg(\frac{2 \pi}{\beta} \bigg)^{N+ \frac{1}{2}}  \exp \bigg( \frac{\beta}{2L_1L_2L_3} \sum_{k=-N}^N \bigg\vert \int d^3 x' \; \chi (\vec{x}',y) e^{-i \vec{p}_k \cdot \vec{x}'} \bigg\vert^2 \bigg) \bigg] \times \nonumber \eeq
\beq \times \xi (\vec{x},y', t- \delta t)  \; \exp \bigg[ i \bigg( \frac{1}{2} \sum_{k_1=-N}^N \sum_{k_2=-N}^N \bigg( T(\vec{p}_{k_1} + \vec{p}_{k_2} = \vec{0})  \times \nonumber \eeq
\beq \times  \bigg( \int d^3 x' \; \frac{\chi (\vec{x}'',y) - \chi (\vec{x}'',y')}{\delta t} e^{-i \vec{p}_1 \cdot \vec{x}''}  \bigg) \bigg( \int d^3 x''' \; \frac{\chi (\vec{x}''',y) - \chi (\vec{x}''',y')}{\delta t} e^{-i \vec{p}_2 \cdot \vec{x}'''}  \bigg) \bigg) - \nonumber \eeq
\beq - \frac{m^2}{2} \sum_{k_1=-N}^N \sum_{k_2=-N}^N \bigg( T(\vec{p}_{k_1} + \vec{p}_{k_2} = \vec{0})  \bigg( \int d^3 x' \;  \chi (\vec{x}',y)  e^{-i \vec{p}_{k_1} \cdot \vec{x}'}  \bigg) \bigg( \int d^3 x'' \; \chi (\vec{x}'',y) e^{-i \vec{p}_{k_2} \cdot \vec{x}''}  \bigg) \bigg) - \nonumber \eeq
\beq  - \frac{\lambda}{4} \bigg(\sum_{k_1=-N}^N \sum_{k_2=-N}^N \sum_{k_3=-N}^N \sum_{k_4=-N}^N \bigg( T (\vec{k_1} + \vec{k_2} + \vec{k_3} + \vec{k_4} = \vec{0})  \times \nonumber \eeq
\beq \times \bigg( \int d^3 x' \; \chi (\vec{x}',y) e^{-i \vec{p}_{k_1} \cdot \vec{x}'} \bigg) \bigg( \int d^3 x'' \; \chi (\vec{x}'',y) e^{-i \vec{p}_{k_2} \cdot \vec{x}''} \bigg) \times \nonumber \eeq 
\beq \times \bigg( \int d^3 x''' \; \chi (\vec{x}''',y) e^{-i \vec{p}_{k_3} \cdot \vec{x}'''} \bigg) \times \bigg( \int d^3 x'''' \; \chi (\vec{x}'''',y) e^{-i \vec{p}_{k_4} \cdot \vec{x}''''} \bigg) \bigg) \bigg)^4  \bigg) \bigg] \bigg\} \label{StepFinal}  \eeq 
If we would like to convert it to continuum equation we can use the following tactic: the equation of the form 
\beq \xi (\vec{x}, y, t) = f (\xi; \vec{x},y,t- \delta t) \eeq
can be generated through the continuum equation 
\beq \frac{\partial \xi}{\partial t} = \frac{1}{\delta} (- \xi (\vec{x},y,t) + f (\xi; \vec{x},y, t)) \eeq
where we have replaced $\delta t$ with $\delta$ in order to make it clear that we are dealing with continuus process, where $\delta$ is merely a constant of nature, as opposed to step by step process with time interval $\delta t$. Thus, we read off 
\beq \frac{\partial \xi}{\partial t} = \frac{1}{\delta} \Bigg\{ - \xi (\vec{x},y,t) + \frac{1}{L_5} \int dy' \;  \bigg\{ \bigg[  \bigg(\frac{2 \pi}{\beta} \bigg)^{N+ \frac{1}{2}} \times  \nonumber \end{equation}
\begin{equation}\exp \bigg( \frac{\beta}{2L_1L_2L_3} \sum_{k=-N}^N  \bigg\vert \int d^3 x' \; \chi (\vec{x}',y) e^{-i \vec{p}_k \cdot \vec{x}'} \bigg\vert^2 \bigg) \bigg]  \xi (\vec{x},y', t)  \; \exp \bigg[ i \bigg( \frac{1}{2} \sum_{k_1=-N}^N \sum_{k_2=-N}^N \nonumber \eeq
\beq \bigg( T(\vec{p}_{k_1} + \vec{p}_{k_2} = \vec{0})  \times  \bigg( \int d^3 x'' \; \frac{\chi (\vec{x}'',y) - \chi (\vec{x}'',y')}{\delta} e^{-i \vec{p}_1 \cdot \vec{x}''}  \bigg) \bigg( \int d^3 x''' \; \frac{\chi (\vec{x}''',y) - \chi (\vec{x}''',y')}{\delta} e^{-i \vec{p}_2 \cdot \vec{x}'''}  \bigg) \bigg) - \nonumber \eeq
\beq - \frac{m^2}{2} \sum_{k_1=-N}^N \sum_{k_2=-N}^N \bigg( T(\vec{p}_{k_1} + \vec{p}_{k_2} = \vec{0})  \bigg( \int d^3 x' \;  \chi (\vec{x}',y)  e^{-i \vec{p}_{k_1} \cdot \vec{x}'}  \bigg) \bigg( \int d^3 x'' \; \chi (\vec{x}'',y) e^{-i \vec{p}_{k_2} \cdot \vec{x}''}  \bigg) \bigg) - \nonumber \eeq
\beq  - \frac{\lambda}{4} \bigg(\sum_{k_1=-N}^N \sum_{k_2=-N}^N \sum_{k_3=-N}^N \sum_{k_4=-N}^N \bigg( T (\vec{k_1} + \vec{k_2} + \vec{k_3} + \vec{k_4} = \vec{0})  \times \nonumber \eeq
\beq \times \bigg( \int d^3 x' \; \chi (\vec{x}',y) e^{-i \vec{p}_{k_1} \cdot \vec{x}'} \bigg) \bigg( \int d^3 x'' \; \chi (\vec{x}'',y) e^{-i \vec{p}_{k_2} \cdot \vec{x}''} \bigg) \times \nonumber \eeq 
\beq \times \bigg( \int d^3 x'''' \; \chi (\vec{x}'''',y) e^{-i \vec{p}_{k_3} \cdot \vec{x}''''} \bigg) \times \bigg( \int d^3 x''''' \; \chi (\vec{x}''''',y) e^{-i \vec{p}_{k_4} \cdot \vec{x}'''''} \bigg) \bigg) \bigg)^4  \bigg) \bigg] \bigg\} \Bigg\}  \label{DynamicsFinal} \eeq 
Finally, the conditions under the sum signs can always be changed to accommodate what we would expect from loop diagrams (where the second order loops would have higher momenta than first order loops if we take the notion of UV cutoff literally), in which case it would no longer match $\phi^{(N)}$ (rather it might be some combination of different $N$-s depending on what restrictions we chose) but it would still be equally well defined theory. 

\subsection*{8. Conclusion}

In this paper we have shown how arbitrary multiparticle state can be described as a pair of two classical fields, $\chi$ and $\xi$, living in ordinary space with a single extra dimension. The field $\chi$ is a hidden variable field that has nothing to do with actual state and, instead, has to do with determining the coarse graining. On the other hand, $\xi$ indeed describes the physical state per Eq \ref{xiGeneralFinal}. Furthermore, creation and annihilation operators were described as taking one ordinary function to the other ordinary function (see Eq \ref{AnnihilateFinal}, \ref{CreateFinal}, \ref{Annihilate0Final}, \ref{Create0Final}). Furthermore, an ensemble of states is also described by one single wave function, as given in Eq \ref{DensityFinal}. In other words we were able to both overcome the problem of many particles as well as ensemble of states (despit the fact that these are very different issues), and describe both in terms of single wave function in ordinary space. 

Apart from that, we provided a choice of two hypothetical rules of dynamics of $\xi (\vec{x}, t)$: namely, Eq \ref{SimplerDynamics} and Eq \ref{DynamicsFinal}. Both of these equations take"classical" form in a sense that they pertain to $\xi (\vec{x},t)$, yet are non-local. The argument in favor of Eq \ref{SimplerDynamics} is that it is considerably simpler, and the argument in favor of Eq \ref{DynamicsFinal} is that it matches quantum states, as defined in Chapters 3-5, more closely. However, since approximate match can't be empirically falsified, I, personally, favor the Eq \ref{SimplerDynamics}.  In both cases, the quantum states in question obey some version of coarse grained QFT. In one case, we reproduce that particular coarse-graining exaclty, in the other case we reproduce it approximately. But, in both cases, the match with conventional states (that are not coarse grained) is always approximate. 

Our approach was based on coarse graining. In future, it could be made more precise by means of space filling curve constructions given in \cite{SpaceFillingCurve1}, \cite{SpaceFillingCurve2} and \cite{SpaceFillingCurve3}. However, even if we did do that, we would still have to cut off the momentum since said constructions work only in finitely many dimensions. And, since we are accepting the fact that QFT is not precise in one way, we might as well accept that it is not precise in some other way as well -- particularly since the random curve that fills the space up to some coarse graining is a lot more natural than carefully designed curve proposed in \cite{SpaceFillingCurve1}, \cite{SpaceFillingCurve2} and \cite{SpaceFillingCurve3}. Nevertheless, it might be interesting to investigate the latter for the future project just to see whether or not we will be able to make rigourous some of the statements that were more hand waving in this paper. 

One weakness of our approach is Ocam's razor, combined with the fact that no new predictions are made. After all, we do not explaine collapse of wave function: we simply re-define quantum states and then existing collapse models would have to be readjusted. This being the case, a lot of people might not like that the equations look a lot more complicated and unnatural than their conventional counterparts if the predictions are identical. From my point of view, however, the important change is ontology, which I view to be worth it as end onto itself since that is what I view as a key difference between quantum and classical, as opposed to anything else. Another objection the reader might have is how do I know that the proposed model is what takes place in nature as opposed to some other, equally complicated yet different, construction? The answer is I don't know. But what I am set to show is that there is no reason to claim that classical logic doesn't work in quantum mechanics; so I gave a counter-example as to how classical logic "might" work, as given in this paper. Of course the reader can think of other counter-examples, but that will only strengthen my point.

\end{document}